 \documentclass[10pt,journal,compsoc]{IEEEtran}

\ifCLASSOPTIONcompsoc
  \usepackage[nocompress]{cite}
\else
  \usepackage{cite}
\fi

\usepackage{booktabs} 
\usepackage{fancyvrb}
\usepackage{listings}
\usepackage{color}
\usepackage{balance}
\usepackage{fancybox}
\usepackage{xspace}
\usepackage{subcaption}
\usepackage{multirow}
\usepackage{url}
\usepackage{array}
\usepackage{graphicx}
\usepackage{bbding}
\usepackage{wasysym}
\usepackage{amssymb}
\usepackage{pdflscape}
\usepackage{tabularx,colortbl}
\usepackage{dirtree}
\usepackage{rotating}

\usepackage{comment}

\usepackage{algorithm}
\usepackage[noend]{algpseudocode}
\usepackage{enumitem}

\usepackage{colortbl}
\usepackage{tabu}
\usepackage[skins]{tcolorbox}
\usepackage{cleveref}

\usepackage{balance}

\usepackage{float}

\floatstyle{ruled}
\newfloat{program}{thp}{lop}
\floatname{program}{Program}

\usepackage{array, booktabs, boldline} %

\ifCLASSINFOpdf
\else
\fi

\definecolor{grayboxcolor}{HTML}{f2f2f2}
\newcommand{\conclusion}[1]{%
	\begin{center}\noindent\thicklines\setlength{\fboxsep}{8pt}\fcolorbox{black}{grayboxcolor}{\begin{minipage}{3.3in}\textbf{#1}\end{minipage}}\end{center}}

\newcommand{\rqi}{RQ1: How prevalent are JavaScript dependency smells?}
\newcommand{\rqii}{RQ2: How do developers perceive dependency smells and their negative impact?}
\newcommand{\rqiii}{RQ3: Why are these smells introduced in JavaScript projects?}

\definecolor{cadmiumgreen}{rgb}{0.0, 0.42, 0.24}

\def\commentsenabled{}

\ifdefined\commentsenabled
\newcommand{\diego}[1]{\textcolor{cadmiumgreen}{Diego: #1}} 
\newcommand{\rabe}[1]{\textcolor{orange}{Rabe: #1}} 
\newcommand{\abbas}[1]{\textcolor{red}{Abbas: #1}} 
\newcommand{\todo}[1]{\textcolor{red}{TODO: #1}}
\else
\newcommand{\diego}[1]{} 
\newcommand{\rabe}[1]{} 
\newcommand{\abbas}[1]{} 
\newcommand{\todo}[1]{} %
\fi

\newcommand{\rev}[1]{{#1}}

\newcommand{\npm}{\textit{npm}\xspace}

\begin{document}
%
\title{Dependency Smells in JavaScript Projects}

%
%
%
%

%

\author{Abbas~Javan~Jafari,
	Diego~Elias~Costa,
	Rabe~Abdalkareem,
        Emad~Shihab,~\IEEEmembership{Senior Member,~IEEE,}
        Nikolaos~Tsantalis~\IEEEmembership{Senior Member,~IEEE}
\IEEEcompsocitemizethanks{\IEEEcompsocthanksitem Abbas~Javan~Jafari, Diego~Elias~Costa and Emad Shihab are with the Data-driven Analysis of Software (DAS) Lab at the Department of Computer Science and Software Engineering, Concordia University, Montreal, Canada.\protect\\
E-mail: {a\_javanj@encs.concordia.ca, diego.costa@concordia.ca, emad.shihab@concordia.ca}
\IEEEcompsocthanksitem Rabe Abdalkareem is with the School of Computer Science at Carleton University, Canada\protect\\
E-mail: rabe.abdalkareem@carleton.ca
\IEEEcompsocthanksitem Nikolaos Tsantalis is with the Department of Computer Science and Software Engineering, Concordia University, Montreal, Canada. \protect\\
E-mail: {nikolaos.tsantalis}@concordia.ca
}
\thanks{Manuscript received October 27, 2020; revised May 06, 2021.}}

\IEEEtitleabstractindextext{%
\begin{abstract}
Dependency management in modern software development poses many challenges for developers who wish to stay up to date with the latest features and fixes whilst ensuring backwards compatibility. Project maintainers have opted for varied, and sometimes conflicting, approaches for maintaining their dependencies. Opting for unsuitable approaches can introduce bugs and vulnerabilities into the project, introduce breaking changes, cause extraneous installations, and reduce dependency understandability, making it harder for others to contribute effectively.

In this paper, we empirically examine evidence of recurring dependency management issues (dependency smells). We look at the commit data for a dataset of 1,146 active JavaScript repositories to catalog, quantify and understand dependency smells.
Through a series of surveys with practitioners, we identify and quantify seven dependency smells with varying degrees of popularity and investigate why they are introduced throughout project history. Our findings indicate that dependency smells are prevalent in JavaScript projects \rev{with two or more distinct smells appearing in 80\% of the projects, but they generally infect a minority of a project's dependencies.} Our observations show that the number of dependency smells tend to increase over time. Practitioners agree that dependency smells bring about many problems including security threats, bugs, dependency breakage, runtime errors, and other maintenance issues. These smells are generally introduced as developers react to dependency misbehaviour and the shortcomings of the \npm ecosystem.
\end{abstract}

}

\maketitle

\IEEEdisplaynontitleabstractindextext

%
\IEEEpeerreviewmaketitle

\section{Introduction}
\label{sec:Introduction}

\IEEEPARstart{S}{oftware} ecosystems have completely changed the way we build software, by enabling code reuse in large scale through software packages. 
Developers now rely on an increasingly high number of packages to build their programs, reusing code to increase productivity, improve software quality and decrease time-to-market~\cite{lim1994effects,mohagheghi2004empirical}.  

However, this development paradigm creates a lot of dependencies, and managing these dependencies has become a key issue for developers~\cite{bogart2016break, artho2012software, decan2019empirical}.
An example incident in \npm (Node Package Manager) is the release of a backward incompatible minor version 1.7.0 of the package ``underscore'' that caused many complaints among dependent packages about underscore not respecting Semantic Versioning (SemVer) \cite{underscore2014}. Another anecdote is the removal of the ``left-pad'' package that caused widespread breakage among big internet sites like Facebook, AirBnB, and Netflix \cite{leftpad2018}.

Semantic Versioning (SemVer) has been presented as a solution to help effectively manage dependencies \cite{semver2019}. It allows maintainers to automatically receive fixes and minor updates, while also limiting their exposure to breaking changes. However, previous research has shown that developers do not \rev{always} conform to SemVer~\cite{kula2018, chinthanet2019lag, dietrich2019dependency, wittern2016look}. This has created major problems due to outdated dependencies and breaking changes. There is evidence that up to 40\% of the packages on \npm rely on at least one package with a publicly disclosed vulnerability~\cite{zimmermann2019small} and up to 53\% of package releases on \npm suffer from some sort of technical lag \cite{decan2018evolution}. Simply allowing dependencies to automatically update is also problematic. A survey of 2,000 developers across different ecosystems has reported that 70\% of the Node.JS developers have experienced breaking changes caused from updates when building their package~\cite{values2020}. 

\rev{Although the reasons and impacts regarding the circumvention of guidelines such as SemVer and opting for alternative approaches} have been referenced in the literature, they are usually studied as a side-topic to explore other issues such as technical lag~\cite{decan2018evolution} or security vulnerabilities~\cite{zimmermann2019small}. 

We argue that in order to better improve the management of dependencies, there is a need for a study that specifically focuses on dependency issues. Hence, the objective of this paper is to catalog, quantify, and understand these dependency issues, which we refer to as dependency smells. \textit{Dependency smells are recurring violations of dependency management guidelines that have negative consequences on the project and the ecosystem}. 

Through our empirical study, we curated and analyzed a dataset of 1,146 open source JavaScript projects. First, we provide the definition and description of seven identified dependency smells: pinned dependency, URL dependency, restrictive constraint, permissive constraint, no package-lock, unused dependency, and missing dependency. 
These definitions are validated with an initial survey of twelve practitioners. We then qualitatively investigate their advantages and disadvantages through a survey with 41 JavaScript practitioners.
We then conduct an empirical study to examine the prevalence of these dependency smells, and investigate why they are introduced in the studied JavaScript projects. Our study is formalized through the following research questions:

\textit{\rqi} We built a tool that detects the aforementioned defined dependency smells in the 1,146 JavaScript projects in our dataset. Our findings reveal that dependency smells are prevalent in JavaScript projects. \rev{The results reveal that 80\% of the projects are infected with two or more distinct smells. Not all smells occur at a large degree. Four out of seven appear in less than 30\% of projects and the majority of smells infect a minority (17\%) of a project's dependencies. However, the results hint at inadequate attention to dependency maintenance or a lack of awareness regarding best practices in dependency management.}

\textit{\rqii} 
We crafted a questionnaire and surveyed 41 practitioners to quantify their agreement/disagreement on the consequences and potential rationales of dependency smells. Developers confirmed the harmful nature of these smells and they were even more critical than we anticipated.

\textit{\rqiii} We asked developers why they opted to introduce a smell \rev{rather than using the alternatives suggested by \npm or SemVer}. We aggregated the 28 responses into 14 reasons. Our findings show that experiencing breaking changes and needing a fix not yet published on \npm were among the most cited reasons for introducing a dependency smell.

\color{black}

Looking through the evolution of the dependency smells over time, we observe that these dependency smells are addressed, but new smells are introduced more frequently than old smells are being fixed. This has caused an overall upward trend in their accumulation.

This study presents (i) A catalog of dependency smells crafted and validated, by quantifying responses from JavaScript practitioners, (ii) A large-scale empirical study of dependency smells in 1,146 popular JavaScript projects, and (iii) A prototype tool named DependencySniffer \cite{dependencysniffer} that can analyze any JavaScript project that uses \npm and detect the presence of dependency smells. The tool can potentially be incorporated into CI pipelines to prevent dependency smells from rippling through software projects. 

The rest of the paper is organized as follows. Section 2 provides a background on dependency management in Javascript, specifically \npm. Section 3 introduces our catalog of dependency smells. Section 4 and 5 present our dataset and the smell detection technique for the empirical analysis. The results for the three RQs are presented in Section 6. We discuss smell evolution in Section 7 and we present our implications in Section 8. The related works are discussed in Section 9. Section 10 highlight the limitations of our study and Section 11 concludes the paper.

\section{Background}
\label{sec:Background}
In this section, we explain the necessary background required to understand our work on dependency smells. We explain how dependency management works in JavaScript using \npm. We then explain the semantic versioning standard and how and why it may not be followed. 

\subsection{JavaScript Dependency Management}

\begin{figure}[h]
  \centering
  \fbox{\includegraphics[width=0.97\linewidth]{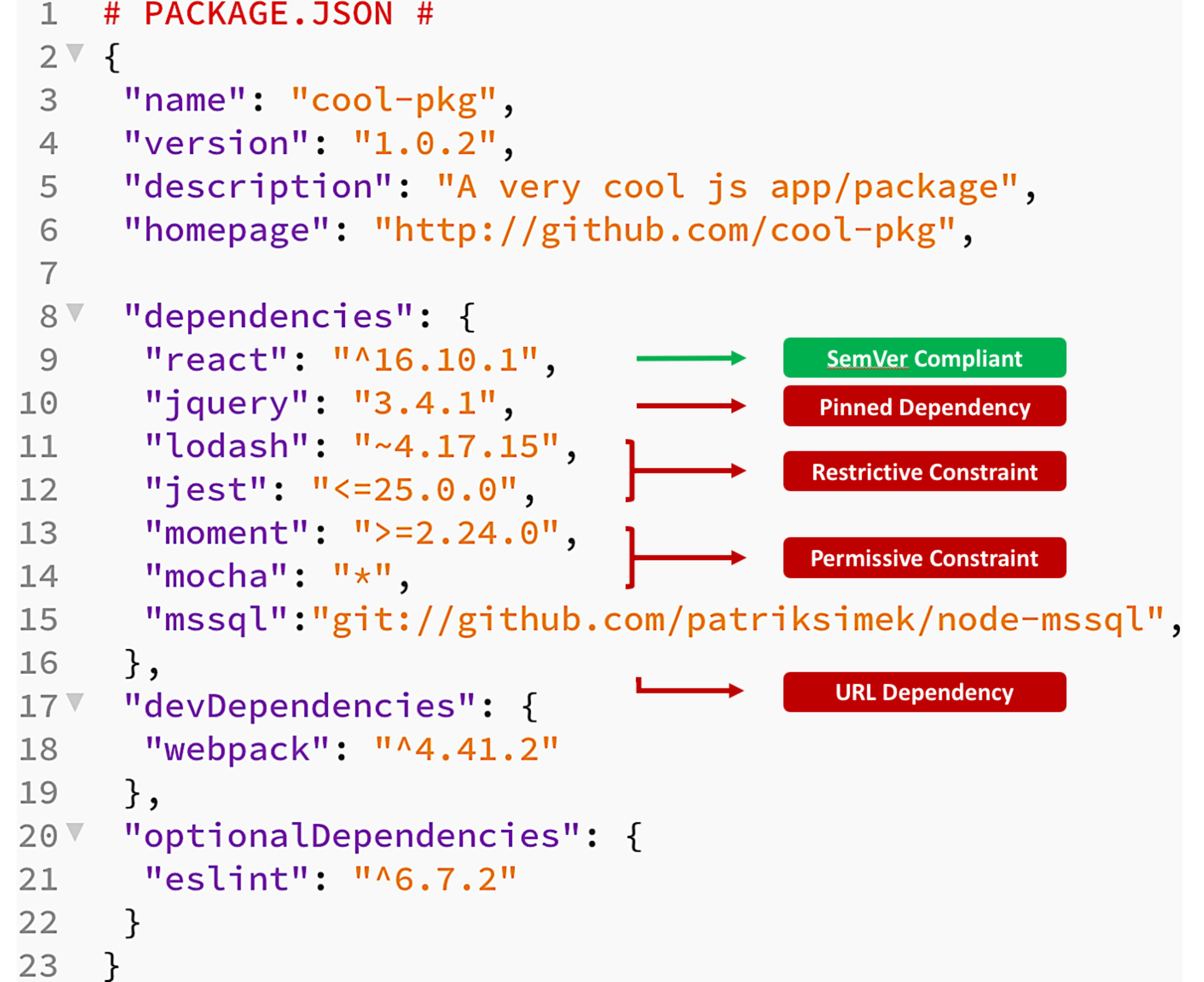}}
  \caption{Example of a smelly package.json file}
  \label{fig:packagejson}
\end{figure}

JavaScript projects generally use the package.json file to specify their package metadata and dependencies \cite{npmpkg2019}. 
This is the de-facto file for determining how a project fits within the \npm dependency network, what types of dependencies it uses, and how restrictive or permissive the developers are in regards to their dependencies. Every time a project is installed, \npm creates a package-lock file which includes the fixed versions of all direct and indirect dependencies used so future installations remain consistent. Figure~\ref{fig:packagejson} presents a generated example of a package.json file.

Developers can define three types of dependencies in package.json: runtime (default), development, and optional dependencies. The runtime dependencies present a list of depended upon packages along with their appropriate versions or version ranges based on SemVer. Development dependencies, unlike runtime dependencies, are only required for development operations such as testing and linters. As for optional dependencies, the \npm package manager will try to fetch them, but if it fails, it will not raise any errors.

\subsection{Semantic Versioning (SemVer)}
Semantic Versioning (SemVer) is considered the de-facto versioning standard for many software ecosystems including \npm and PyPI. It was introduced by the co-founder of GitHub, Tom Preston-Werner, in 2011. In our study, we focus on SemVer 2.0, which was released in 2013 \cite{semver2019}. The purpose of SemVer is twofold. First, it allows for package developers to communicate the extent of backward incompatible changes in their new releases to package dependents. Secondly, it allows for dependents of a package to specify how restrictive or permissive they want to be in automatically accepting new versions of the packages.

SemVer proposes a multi-part versioning scheme in the form of \textbf{major.minor.patch[-tag]} parts, which helps in identifying the type of changes in the newly released package. Any change to the newly release package that is backward incompatible (e.g. break the API) requires a change to the \textbf{major} version. New functionality that does not create backward incompatibility in the newly released package requires a change to the \textbf{minor} version. The \textbf{patch}  represents backward compatible fixes (e.g., fixing a bug). There is also an optional \textbf{tag} for specifying pre-releases (e.g. 1.2.3-beta) \rev{and other build metadata such as post-release build numbers}. Developers must respect these guidelines in order to be SemVer compliant \cite{semver2019}.

\subsection{SemVer Non-Compliance}
While SemVer is a promising solution to many dependency update issues, and even though it is recommended by ecosystem maintainers \cite{npmsemver,pipsemver}, it is not always followed in practice. A survey of more than 2,000 developers from 18 different software ecosystems including the \npm ecosystem showed that while 92\% of the respondents for \npm claim to only increment the left-most digit if they release an update that may break dependent code, 70\% of the surveyed developers had a different experience when updating their dependencies \cite{values2020}.
In these cases, they experienced breaking changes even when updating in compliance with SemVer guidelines. This leads developers to exercise caution when following SemVer, especially for dependencies which have a history of non-compliance for their releases. 

\npm uses many different notations to allow developers to precisely specify the desired versions for a dependency. In order to be compliant with SemVer, developers should allow automatic updates for the minor and patch versions for all post-production (1.0.0 or higher) releases.

\begin{itemize}
	\item \npm uses caret (\textsuperscript{$\wedge$}) to accept minor and patch updates but developers can use constraints in the form of 2.x. 
	\item \npm uses tilde ($\sim$) to accept patch releases only, but developers can use constraints in the form of 2.1.x.
	\item \npm also allows the use of * or " " wildcards to accept all future versions but developers can also use constraints in the form of $>$=2.1.0 which only limits the lower bound of the dependency.
	\item It is possible for developers to only use a specific version of a dependency and refuse automatic updates. 
	\item \rev{Developers can also opt for other approaches such as using a version range in the form of version1-version2 or specifying a URL to fetch dependencies.}
\end{itemize}
\section{Dependency Smells}
\label{sec:Dependency smells}

In this section, we define each dependency smell, the rationale for why it might occur, and its negative consequences (\Cref{sub:smells-catalog}). \rev{Similar to the code smell literature, whether or not something is a smell is based on the context \cite{fontana2016antipattern}}.
 These smells have not been explicitly defined in the literature. They are initially observed by studying violations of \npm recommendations, violations of SemVer guidelines and by studying the discussions surrounding the package.json file. We have followed up the initial observations with a survey from twelve practitioners in the field to better understand the negative consequences of each smell and why they might occur.
We sent out an open-format questionnaire (where respondents are free to write anything) to seventeen JavaScript developers that we knew had sufficient experience with developing JavaScript projects. \rev{The developers were a convenient sample from Canada and Brazil that were known by the authors. They were contacted by email.} We presented them with different approaches of depending on an \npm package and asked them to write down the advantages and disadvantages associated with each approach. Since we did not want to bias the developers into associating negative consequences with the presented approaches, we refrained from calling an approach a ``smell'' and also included the SemVer-compliant alternative among our approaches. More importantly, we asked them to give us both advantages and disadvantages for each approach. The advantages not only help in reducing bias but also help in understanding why a particular dependency smell might occur. In order to ensure a similar understanding of dependency management and SemVer, we provided the participants with links to the \npm dependency reference \cite{npmpkg2019} and the SemVer standard guide \cite{semver2019} .

From the seventeen invites, twelve developers responded to our first survey (i.e,. 70\% of response rate).
From the 12 respondents, 10 identified themselves as industry practitioners and 2 participants were students.
All participants have used \npm and 11 out of 12 participants had at least one year of experience in Node.JS development. 
From the 12 respondents, 11 were familiar with SemVer.
For each dependency smell, we categorized the 12 responses using an open-coding approach similar to the guidelines expressed by Philip Burnard \cite{burnard1991method}. Through an iterative process for each smell, we extracted and grouped relevant themes from developer responses into reasons for why a smell occurs and reasons for why it can cause problems.

\rev{The smells identified in this paper focus on \textit{how} dependencies are used and managed, rather than \textit{what} dependencies are selected. Therefore, issues such as using too many dependencies or the wrong selection of dependencies are not covered, as they are specific to the requirements and functionalities of each project.}

\subsection{Dependency Smells Catalog}
\label{sub:smells-catalog}

This smells catalog consists of a list of seven identified dependency smells. We do not claim this to be an exhaustive list, but given that \npm limits the possible expressions of dependency constraints in the package.json file, the smells presented here cover the majority of different alternatives for depending on a package. Figure~\ref{fig:packagejson} presents examples of the restrictive constraint, pinned dependency, permissive constraint, and URL dependency smells. Other dependency smells are not identifiable solely by looking at the package.json file and they require closer inspection of the source code or project directory.

In identifying the negative consequences associated with each pattern, along with the reasons for why they might occur, we relied on developers' feedback as well as the inherent \npm rules for evaluating dependency constraints. In each smell's definition, we give examples of developer responses (Tagged P1 through P12). Table~\ref{tab:overview} presents an overview of the smells and their negative consequences.

Note that in this first survey we did not ask about the Unused Dependency and the Missing Dependency smell.
\rev{These smells are included as they are both a clear case of dependency mismanagement. They create a mismatch between dependencies used in the source code and the ones defined in package.json which is a violation of npm guidelines \cite{npmpkg2019}. Hence, we do not expect any developers to consider them as a valid approach with positive outcomes. Missing dependencies can cause runtime errors or crashes when executing the code that relies on the dependency. Unused dependencies unnecessarily increase installation size and maintenance effort required for dependencies. Soto-Valero et al. \cite{soto2020comprehensive} use the term bloated dependencies to study unused dependencies in the Maven ecosystem. Depcheck is an npm tool which reports unused and missing dependencies for npm projects \cite{depcheck2019}.}
Additionally, we did not ask developer’s opinion on not using package lock, which is the smell in our catalog (No package-lock). Instead, we ask the advantages and disadvantages of \textit{using} package.lock in the project, given this more intuitive and easier to assess than the consequences of not using an approach. We then use the pros of using a package.lock as the cons of not having package.lock in the project, and vice versa.
\\
\begin{table*}[t]
	\centering
	\caption{Overview of dependency smells.}
	\label{tab:overview}
	\begin{tabular}{c l l l}
		\toprule
		\centering
		\textbf{S\#} & \textbf{Smell} & \textbf{Description} & \textbf{Consequence}\\
		\midrule
		
		S1 & Pinned Dependency & Using a fixed version of a dependency & Not receiving any fixes and manual update workload \\
		S2 & URL Dependency & Fetching dependency from a specific URL & Increased security risks and link inaccessibility \\
		S3 & Restrictive Constraint & Only updating dependencies to patch releases & Not receiving compatible features and some fixes\\
		S4 & Permissive Constraint & Allowing dependency updates to major releases & Significantly greater risk of breaking changes\\
		S5 & No Package-Lock & Not including package-lock in source repository & No guarantee of consistent installation\\
		S6 & Unused Dependency & Installing unused dependency packages & Unnecessarily bloated dependency folder\\
		S7 & Missing Dependency & Omitting a needed dependency from package.json & Code breakage or ambiguous dependencies\\
		\bottomrule
	\end{tabular}
\end{table*}

\noindent\textbf{\large{S1 - Pinned dependency: }}

\noindent\textbf{Description:}
A pinned dependency is a type of restrictive approach in which only a single version of the dependency is accepted, for example pinning to version 1.2.3.

\noindent\textbf{Why it occurs:}
Choosing to pin dependencies is a common approach to ensure fine-grained control over the software packages that a project depends on. It is sometimes even recommended for runtime dependencies of deployable projects to ensure consistent installations. 
This approach drastically reduces the risk of breaking changes and unknown bugs or vulnerabilities that may exist only in newer releases. For example, P12 said: \textit{``Makes it extremely unlikely for breaking changes to happen ...''}.

\noindent\textbf{When it is a dependency smell:}
By pinning a dependency, we will not receive important security and bug fixes, as P12 mentions: \textit{``you are more exposed to security risks and bugs"}. It therefore leaves our project less and less secure over time, unless manual measures are taken to constantly update the constraint to a newer version, which will require extra effort. For example, P10 said: \textit{``[It] creates technical debt in dependencies. Someone has to put in the time later to update them ...''}. Having pinned dependencies is even worse for packages which will later be used as dependencies in other projects, since \npm can no longer re-use similar dependencies for installed packages but must rather have separate installations for the different versions, which can cause a significant increase in the number of installed dependencies for a project. For example, P10 cited: \textit{``...increases node modules size because different packages can’t share the same installation''}.
\\

\noindent\textbf{\large{S2 - URL dependency: }}

\noindent\textbf{Description:}
URL dependencies are constraints that directly point to an online URL, often pointing to a repository link such as GitHub. The URL may point to a general repository or a specific release or branch in that repository. Figure~\ref{fig:packagejson} contains an example of a URL dependency.

\noindent\textbf{Why it occurs:}
The most plausible reason for choosing a URL dependency is in cases where there is no equivalent on the \npm package manager, as P2 states: \textit{``Point to a package outside \npm listings''}. Another potential reason is instances where the dependent strongly needs a new fix in the repository that has not yet made its way to the latest release on \npm. For example, P7 said: \textit{``... the fix is applied to the GH repo way before we see it on \npm. ...''}.

\noindent\textbf{When it is a dependency smell:}
URL dependency constraints are problematic for a variety of reasons. If the URL points to a master branch of a software repository it can lead to breaking changes since any new change to the master branch will be fetched. If the URL points to a specific release of a package on a repository or other type of webpage, it is akin to a pinned dependency. In any case, it is very difficult to adhere to SemVer guidelines whilst using URL constraints for dependencies. In addition, since developers are operating outside \npm, they cannot take advantage of its project metrics and security advisories, and they are more vulnerable to typosquatting attacks (different and malicious packages with very similar names) \cite{npmtypo2017}. Apart from security issues, the unstable nature of unreleased packages can also introduce more bugs, as mentioned by P9: \textit{``... vulnerable to typo-squatting attacks and even if the repository is legit, you are depending on something unstable ...''}. It can also be harder to identify if a new version is released if the URL does not point to a code repository. Additionally, \npm prohibits the deletion of uploaded packages after a grace period, but a URL can change or be removed completely at the owner's discretion, leaving developers with missing dependencies. For example, P8 cites: \textit{``If the linked library disappear then code will stop working ...''}. If the package exists on \npm, one should use the published version. Otherwise, it may be possible to avoid this dependency in favor of its \npm-established alternative.
\\

\noindent\textbf{\large{S3 - Restrictive constraint: }}

\noindent\textbf{Description:}
Restrictive constraints are any dependency constraints that are more restrictive than SemVer. For post-1.0.0 releases, this is equivalent to only accepting patch updates using the "\textasciitilde" notation (e.g. \textasciitilde2.1.3) or by using the "x" notation in the patch component of a constraint (e.g. 2.1.x).
While pinned dependencies are a special case of restrictive constraint, we opt for separating them in our study for clarity purposes.

\noindent\textbf{Why it occurs:}
When a developer knowingly opts for a more restrictive approach in specifying dependency constraints, it is likely due to fear of breaking changes from automatic updates. For example, P12 said: \textit{``... reduce[s] the chance of breakage"}. Indeed, adhering to SemVer only guarantees backward compatible updates if the maintainers of the dependency also adhere to SemVer. This verification is not straightforward and it can be difficult to ensure backward compatibility in practice \cite{decan2019}. Nonetheless, restrictive constraints still allow patch updates, as P8 states: \textit{``This will make sure your libraries are up to date in regard to security patches"}.

\noindent\textbf{When it is a dependency smell:}
Restrictive constraints prevent the project from receiving backward compatible updates. For example, P10 mentions: \textit{``I will not get "safe" feature updates"}. In practice however, the situation could be even worse. Since minor releases can also contain crucial fixes which are not necessarily back-ported to previous patch releases, using the restrictive constraint does not even guarantee we will receive the majority of important security or bug fixes. For example, P9 states: \textit{``quite common to see lots of bugs and vulnerabilities addressed in a minor which never make it to any of the previous patch releases"}. Alternatively, restrictive dependencies have to be manually changed more frequently to keep up with newer versions. This creates extra work for developers, as mentioned by P2: \textit{``Increased work required to update the system"}.
\\

\noindent\textbf{\large{S4 - Permissive constraint:}}

\noindent\textbf{Description:}
Permissive constraints are any dependency constraints that are more permissive than SemVer. For post-1.0.0 releases, this is equivalent to accepting major updates using the * or "" wildcards, or by specifying a minimum version range without a maximum limit such as $>1.2.3$.

\noindent\textbf{Why it occurs:}
In cases where there is an inter-dependency between dependencies maintained by the same developers, they might freely accept any updates since they are in charge of both projects and are therefore more confident in the type of changes they make. For some, removing restrictions is an easy way to make sure the latest updates are fetched. For example, P9 said: \textit{``It caters to my inner laziness"}.

\noindent\textbf{When it is a dependency smell:}
Permissive constraints expose projects to the constant risk of breaking changes, as stated by P10: \textit{``Sooner or later, an update will break the API"}. If the maintainers pinpoint the breaking changes immediately, they have to manually change the constraint to an earlier version, which they have first to identify to work correctly. This is not always easy if semantically breaking changes have creeped up during several updates. In addition, it is irrational to claim a project will work with all future versions of a dependency. In fact, the Cargo ecosystem has prohibited the use of wildcards for this very reason \cite{cargowild2018}.
\\

\noindent\textbf{\large{S5 - No package-lock: }}

\noindent\noindent\textbf{Description:}
The \npm package-lock file includes the entire dependency tree with precise dependency versions at a specific time, which make sure the project will be built. The \npm documentation recommends that the package-lock should be committed into source repositories \cite{npmpkglock2019}. However, we observe that developers do not always include package-lock (or its similar counterparts: yarn.lock or npm-shrinkwrap.json) in their project repositories.

\noindent\textbf{When it occurs:}
Since package-lock is auto-generated, this smell occurs because developers chose not to commit the file into their repository. Since the package-lock does not affect the package.json file and it can be rebuilt if the desired dependencies change, or it can be ignored altogether, we do not see any rationale for not including it, and developers seem to generally agree. For example, P12 said: \textit{``I don't see any issues with having package-lock"}.

\noindent\textbf{Why it is a dependency smell:}
In the absence of the package-lock file, there is no way to guarantee that installation at different times will yield the same result (even if the dependencies are pinned), as stated by P10: \textit{``everyone can use it [package-lock] to install the same exact thing"}. In a sense, package-lock provides the best of both worlds by allowing developers to comfortably use SemVer in their package.json file while giving the option to install using a pinned dependency tree. Without a package-lock there is no way to track the history of dependencies without uploading the large node\_modules folder. For example, P9 mentions: \textit{``... if I don’t use package-lock ... I need to push my node modules folder"}.
\\

\noindent\textbf{\large{S6 - Unused dependency:}}

\noindent\textbf{Description:}
The unused dependency smell represents runtime dependencies that exist in the package.json file, but are not used in source code of the project.

\noindent\textbf{Why it occurs:}
The unused dependency occurs when the code and import statements that needed the dependency were removed but the package.json file has not been cleaned up to properly reflect this. Alternatively, the dependency may only be used in development environments but it is incorrectly listed under runtime dependencies instead of the development dependencies in the package.json file.

\noindent\textbf{When it is a dependency smell:}
If the dependency was actually removed from the project, then including it in the package.json file will cause an extraneous package installation and take up unnecessary space in the modules folder. Even if the dependency is used in development but incorrectly specified under runtime dependencies rather than development dependencies, it will result in the same extraneous installation when installing a project for production. Additionally, both cases will cause confusion for users and developers of the project. In any case, a mismatch between the dependencies imported in the source code and the ones defined in the package.json is a violation of \npm guidelines \cite{npmpkg2019}.
\\

\noindent\textbf{\large{S7 - Missing Dependency: }}

\noindent\textbf{Description:}
The missing dependency smell represents packages that are used in the source code, but not specified in the runtime dependencies of the package.json file.

\noindent\textbf{Why it occurs:}
The missing dependency smell can occur when a dependency is manually installed but not added to the package.json (by using the `--save' option with the \npm install command). Additionally, it can appear because the dependency is removed from the project, and thus the package.json file, but the import statements are not fully cleaned up from the code. Alternatively, a missing dependency may still be used in the code but it has not yet raised errors because it exists in a seldom-reached execution path.

\noindent\textbf{When it is a dependency smell:}
Missing dependencies can be very problematic if they are in fact used in the source code but are not properly specified in the package.json file. This can result in bugs or crashes while executing the part of the source code that relies on this dependency. If the missing dependency occurs because the dependency is imported in source code but never actually used, it will still create ambiguity for users and developers of the source code. In any case, a mismatch between the dependencies imported in the code and the ones defined in the package.json is a violation of \npm guidelines \cite{npmpkg2019}.

\color{black}

\section{Dataset}
\label{sec:data-curation}

We gather a large set of open-source JavaScript projects that are hosted on GitHub to create our dataset.
We chose to study projects written in JavaScript since it is the most popular programming language on Github~\cite{TheState18online}. We also study \npm since it is the official registry of JavaScript packages with more than 1.5 million packages \cite{libraries.io2020}.

We start our data collection by retrieving the GitHub metadata from GHTorrent~\cite{Ghtorrent2019} containing data until June 1st, 2019.
We use this metadata to identify non-forked open-source Javascript projects that have a minimum of 10 contributing authors (as determined by GitHub) and at least 10 commits since January 2019, that is, six months before the data collection started. 
With this criteria, we aim at filtering projects that are actively maintained by at least a medium sized team of developers and reduce the chances of skewing our results towards abandoned and toy projects, prevalent on GitHub ~\cite{Kalliamvakou:14:MiningGithub,abdalkareem2017}.

Using our selection criteria, we identified a total of 2,216 candidate projects. 
We further prune this dataset by 1) removing 569 projects that do not contain package.json, making it impossible to reliably pinpoint their dependencies and 2) removing 413 projects that had no runtime dependencies in their package.json, as development and optional dependencies are not in the scope of this study.
We were left with 1,146 projects as our dataset. This accounts for 26,924 dependency constraints (in the latest snapshot).

\begin{figure*}
	\centering
	
		\begin{subfigure}{.24\linewidth}
			\includegraphics[width=\linewidth]{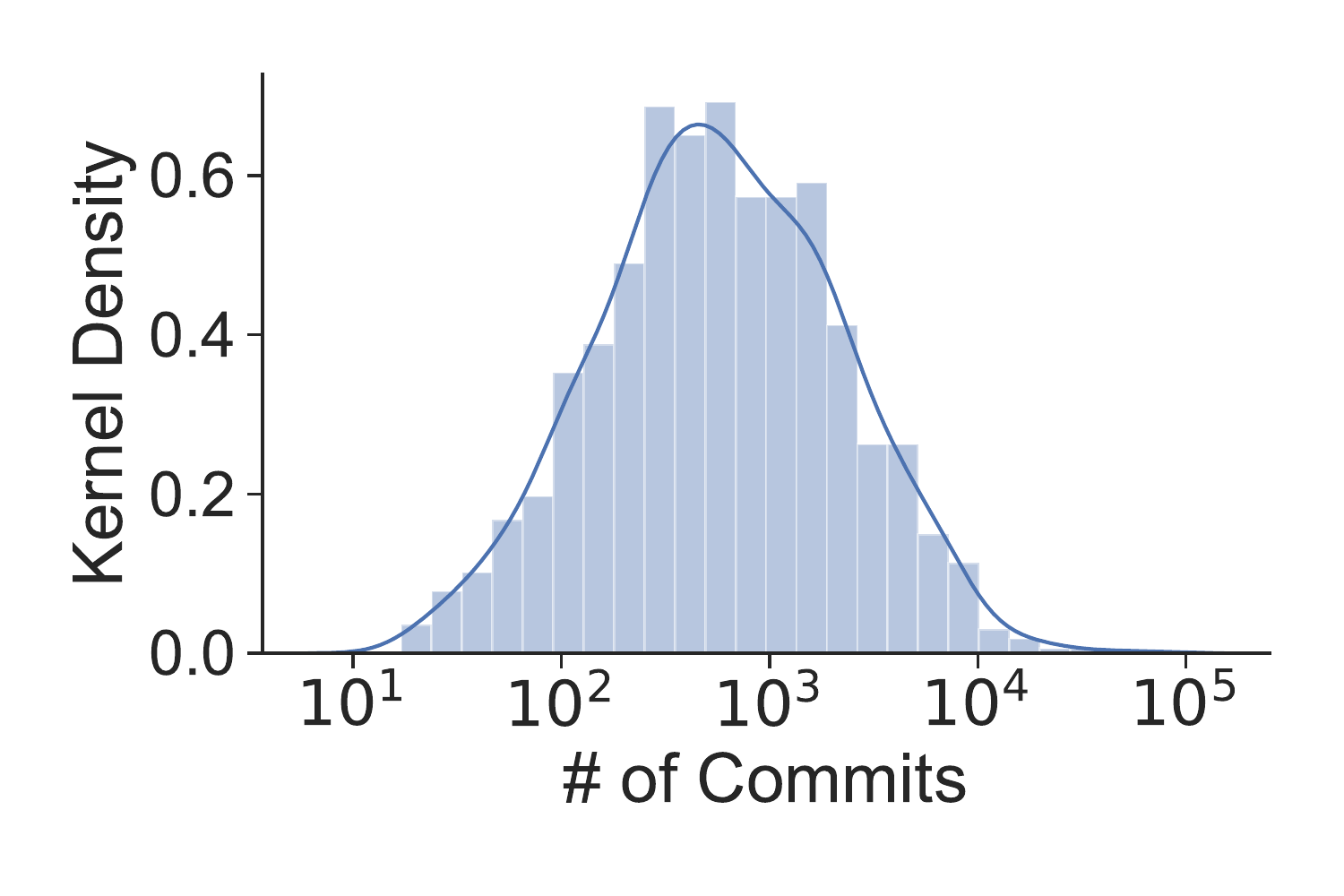}
			\caption{Number of commits}
		\end{subfigure}
		\begin{subfigure}{.24\linewidth}
			\includegraphics[width=\linewidth]{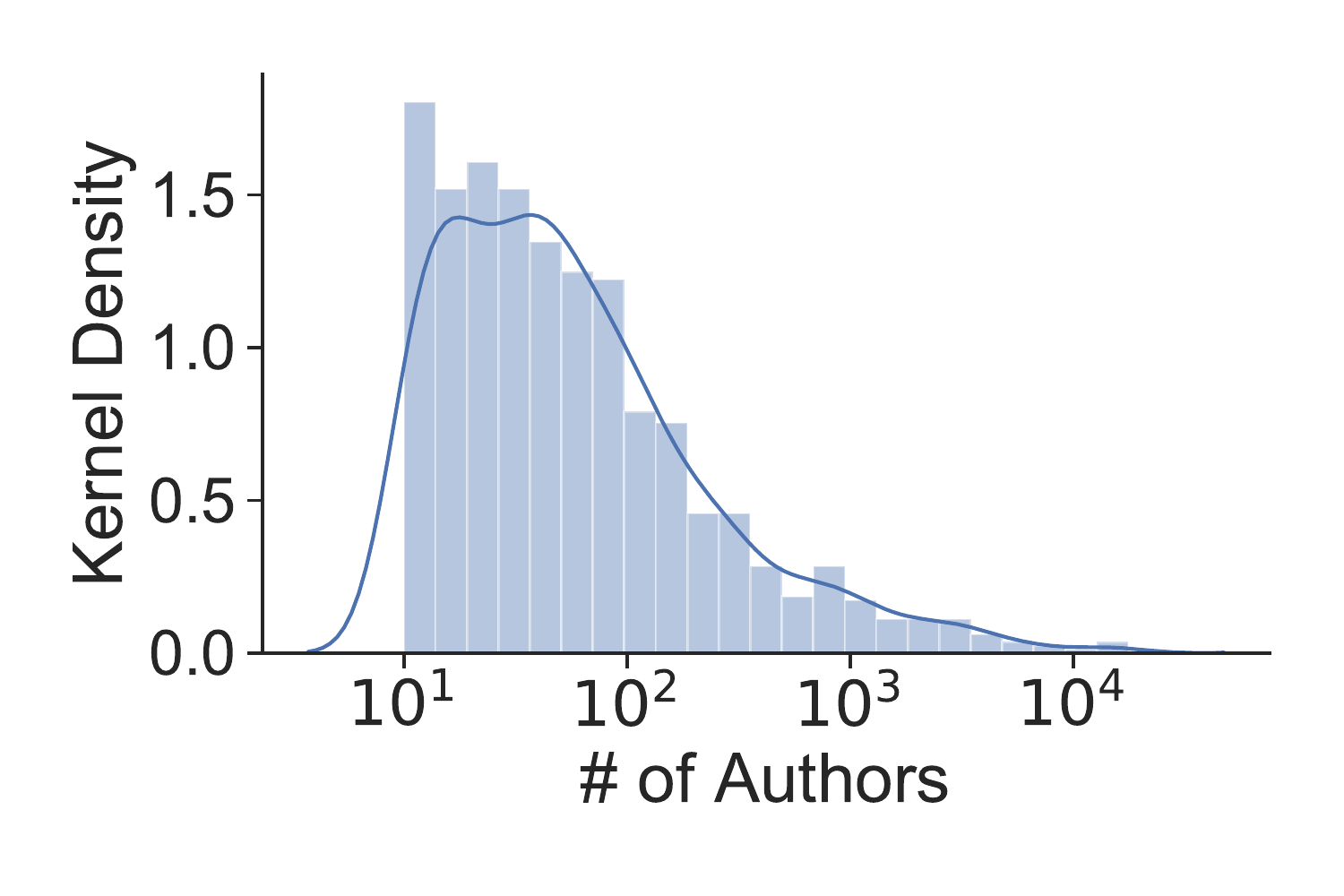}
			\caption{Number of authors}
		\end{subfigure}
		\begin{subfigure}{.24\linewidth}
			\includegraphics[width=\linewidth]{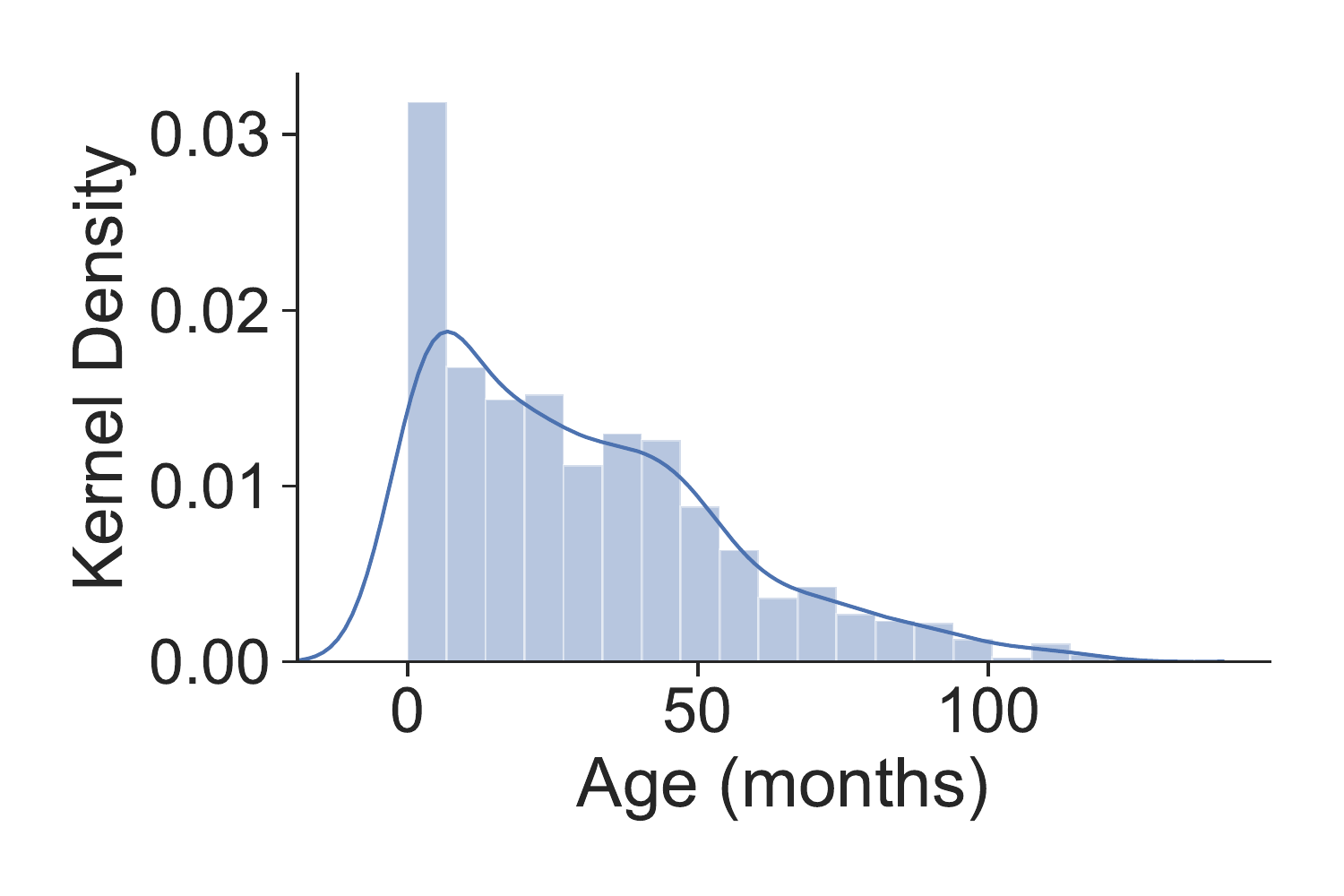}
			\caption{Age of the projects (months)}
		\end{subfigure}
		\begin{subfigure}{.24\linewidth}
			\includegraphics[width=\linewidth]{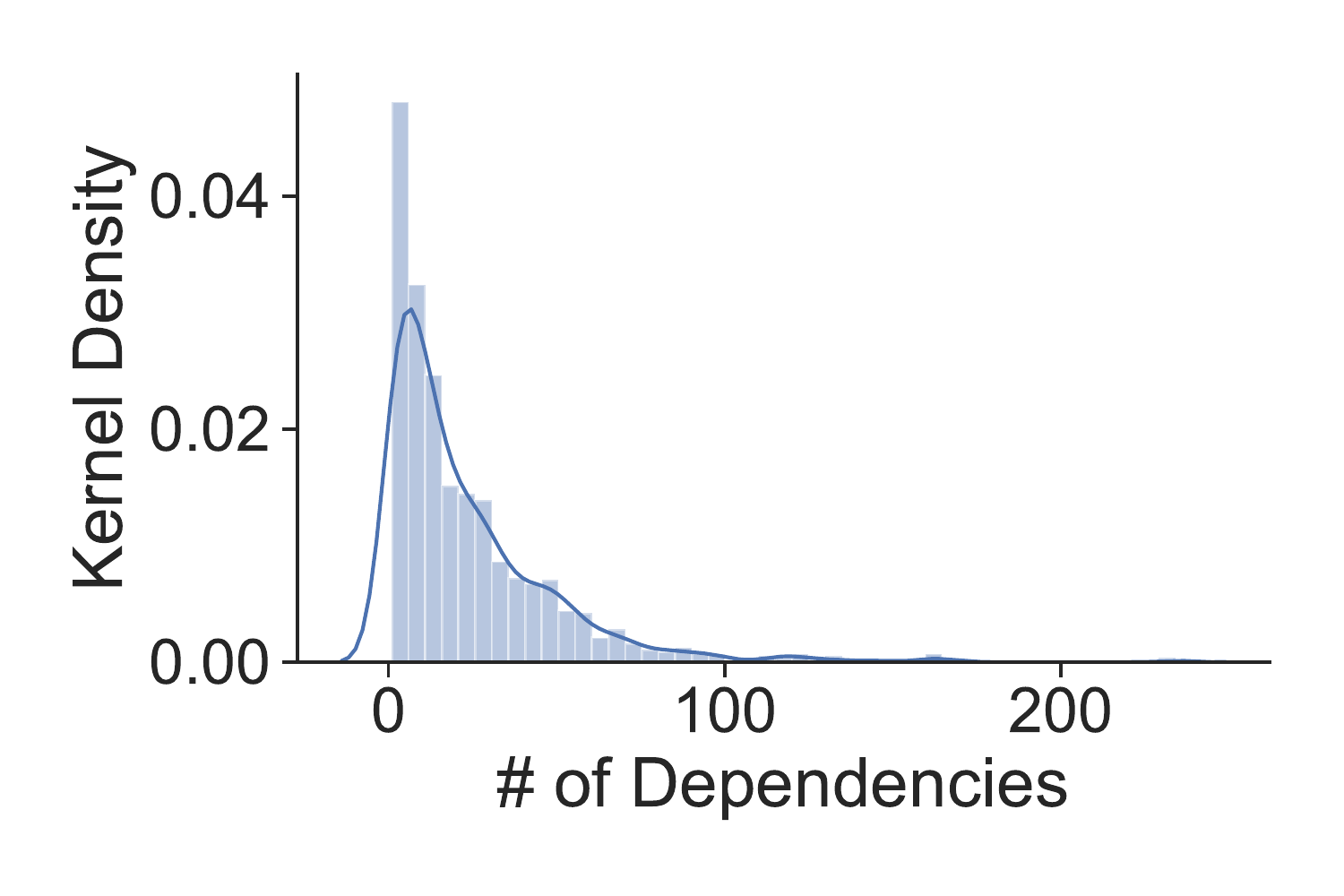}
			\caption{Number of dependencies}
		\end{subfigure}
	
	\caption{{Distribution of projects in the dataset under four perspectives: age, authors, commits and dependencies}} 
	\label{fig:dataset-dist}
\end{figure*}

Figure~\ref{fig:dataset-dist} presents the dataset distribution for the number of commits per project, the number of authors per project, the projects' age in months \rev{and the number of dependencies per project}. 
Our dataset is comprised of 1,541,504 commits, with a median of 555 commits per project. It also contains 29,400 total authors with a median of 21 authors per project. The median age for the projects is 24 months. \rev{The dataset contains a total of 26,924 dependencies with a median of 15 dependencies per project.}
These statistics are indicative of the diversity of our dataset, comprised of well-maintained and active open-source JavaScript projects. \rev{A replication package of our study is available on Zenodo \cite{replication}}.

\section{Smell Detection}
\label{sec:smell-detection}

Since we are aiming at analyzing a large dataset of JavaScript projects, some of our analyses are automated using a tool we developed specifically for this purpose.
Our tool can easily and effectively be used to analyze any JavaScript project and detect current dependency smells. The tool prototype (\texttt{DependencySniffer}) is open source and accessible to everyone \cite{dependencysniffer}. 

For the first four smells (pinned dependency, URL dependency, restrictive constraint, permissive constraint), we go through the projects in our dataset and parse the package.json and their related commits and commit diffs. The parser is a prominent part of our detection because the plain textual diffs are not useful for automated analysis and they can be error-prone. We first parse the package.json in the latest snapshot of the project and store all dependency profiles in a structured SQL database. These dependency profiles can be queried to obtain information such as the name, version constraint and existing smells.
The first four smells are identified using regex rules on the version constraints. We manually checked the results of 50 smell extractions to fine-tune the rules and ensure the rules properly identify the first four smells.
The fifth dependency smell (No-package-lock) can be identified by looking through a projects root directory. We used the depcheck external tool to find unused and missing dependencies. 
In order to investigate the evolution of dependency smells over project history, we need to gather the historical information.
We use \texttt{git log} to identify the commits that modify the package.json file. We then run a large contextual \texttt{git diff} on those commits, i.e., diffs that also include unchanged lines, to properly parse the entire JSON dependency object into a SQL database.
To do so, we perform an intersection of the dependencies in the \textit{before} and \textit{after} commits. The elements that are not in the intersection from the before commit are "deleted" dependencies. The elements that are not in the intersection from the after commit are "added" dependencies. The elements in both the "added" and "deleted" sets that refer to the same dependency but with a different version or different constraint are "modified" dependencies.
This database contains added, removed, and modified dependencies in each commit.
Differentiating between these operations are important because we do not want to flag a smells as fixed, due to the removal of the dependency from the package.json file. 
Aside from the dependency profile, this database captures commit-relevant information (e.g., commit message, commit date) needed for answering RQ3. 
\section{Results}
\label{sec:Results}
In this section, we motivate our research questions, describe the approaches to answer them, and present their results.
\subsection{\rqi}
\label{sub:rq1}

\noindent\textbf{Motivation:}
Thus far, we have catalogued seven dependency smells and described their negative consequences in software projects. 
In this question, we aim at finding empirical evidence of their existence to better understand their prevalence and importance. We look at the current snapshot of projects to see which smells are more common, which will give us a clear picture of the current landscape and guide our in-depth analysis.

\noindent\textbf{Approach:}
We parse the package.json file of each project in our dataset and create a structured database for the dependencies (Section~\ref{sec:smell-detection}) to identify the occurrence of the dependency smells S1 to S4 (see Table~\ref{tab:smells}). 
For the no package-lock smell (S5), we verify the existence of package-lock.json and other lock files including npm-shrinkwrap.json and yarn.lock in the root directory of the studied projects.
To identify the Unused Dependency (S6) and Missing Dependency (S7) smells, we resort to the Depcheck tool~\cite{depcheck2019}.
Depcheck is a tool for analyzing dependencies in projects, reporting instances of unused dependencies (i.e., mentioned in package.json but not used in the code) or missing dependencies (i.e., absent in package.json but used in the runtime source code). It works by analyzing the package.json and comparing the declared dependencies with the dependencies used in the code. We parse Depcheck's output to analyze the occurrence of the two dependency smells.

\begin{table}
	\centering
	\caption{Dependency smells in extracted projects.}
	\label{tab:smells}
		\begin{tabular}{clrr}
		\toprule
		\textbf{\#} & \textbf{Dependency Smell} & \multicolumn{2}{c}{\textbf{Projects (\%)}}\\
		\midrule
		S1 & Pinned dependency & 598 & (52.2\%)\\
		S2 & URL dependency & 117 & (10.2\%)\\
		S3 & Restrictive constraint & 106 & (9.2\%)\\
		S4 & Permissive constraint & 41 & (3.6\%)\\
		S5 & No package-lock & 304 & (26.5\%)\\
		S6 & Unused dependency & 915 & (79.8\%)\\
		S7 & Missing Dependency & 729 & (63.6\%)\\
		\bottomrule
	\end{tabular}
\end{table}

\begin{figure*}
	\centering
	\includegraphics[width=\linewidth]{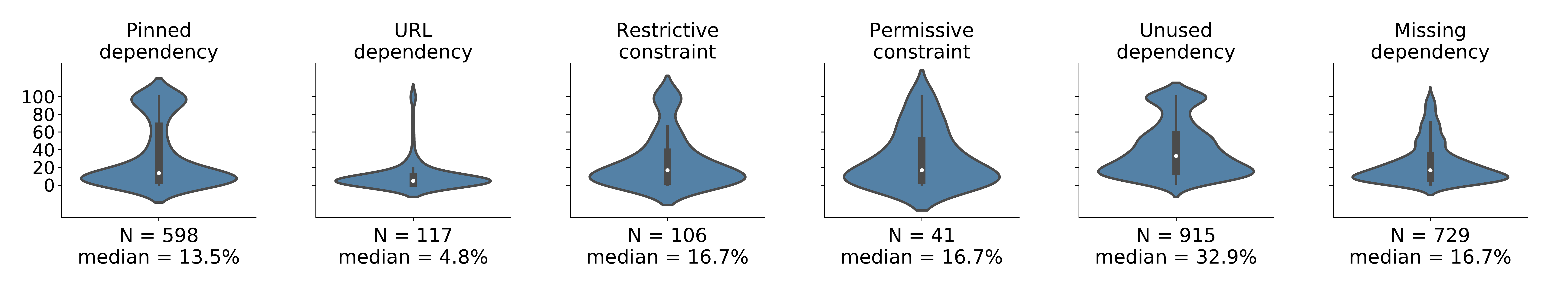}
	\caption{Distribution of dependency smells ratio in the latest snapshot of projects that contain at least one instance of the smell. We specify the number of projects plotted (N) and the median of the distribution (median). } 
	\label{fig:dist-smells}
\end{figure*}

\noindent\textbf{Results:} Table~\ref{tab:smells} shows the dependency smells and the number and  percentage of JavaScript projects in our study that have these dependency smells.
Overall, our findings show that dependency smells are prevalent, with four out of seven appearing in more than 25\% of the projects. As can be seen in Table~\ref{tab:smells}, developers choose to pin some of their dependencies in over 52\% of the projects. This is because while pinning dependencies has many drawbacks, it is still common practice among developers who do not trust their depended-upon packages to properly abide by SemVer.

\begin{table}
	\centering
	\caption{Number of projects containing distinct smell types}
	\label{tab:smell_occurrence}
		\begin{tabular}{lrr}
		\toprule
   		\textbf{Number of Distinct Smells} & \textbf{\# of Projects} & \textbf{\% of Projects} \\
		\midrule
		Zero & 41 & 3.6\% \\
		One & 188 & 16.4\%\\
		Two & 362 & 31.6\%\\
		Three & 357 & 31.2\%\\
		Four & 160 & 14\%\\
		Five & 32 & 2.8\%\\
		Six & 6 & 0.5\%\\
		Seven & 0 & 0\%\\
		\bottomrule
	\end{tabular}
	
\end{table}

The package-lock could be used as a more appropriate alternative for pinning (although not exactly the same thing), since it pins the entire dependency tree while also allowing developers to abide by SemVer. \rev{Even if the project is not meant to be directly installed, it aides development cycles by preventing breakage in transitive dependencies (Section 3 presents other reasons for using package-lock). In any case, this lock file is effortless to generate and will be ignored when the package is fetched as a dependency from \npm.
However, package-lock is not used in over 26\% of the projects.}

In order to investigate this further, we cross-reference the data to identify whether these two dependency smells occur in the same projects, and present this analysis in Table~\ref{tab:co-occur}.
The results are quite interesting. More than 40\% of the studied projects contain both package-lock and pinned dependencies (row 1 in Table~\ref{tab:co-occur}).
We also observe that more than 68\% of the projects that have a package-lock did not feel the need to pin all of their dependencies (row 2 of Table~\ref{tab:co-occur}). Conversely, 11.8\% of the projects had at last one instance of a pinned dependency but did not opt to include the package-lock. Note that the percentages in Table~\ref{tab:co-occur} should not add up to 100\% because a single project can contain both pinned and unpinned dependencies.

\begin{table}
\centering
  \caption{Co-occurrence of pinned constraints with the existence of package-lock.}
  \label{tab:co-occur}
  \begin{tabular}{llr}
     \toprule
   	\textbf{Package-lock} & \textbf{Dependency smell}  & \textbf{Projects (\%)}\\
    \midrule
    Exists & Contains pinned dep. & 463 (40.4\%)\\
    Exists & Contains unpinned dep.  & 783 (68.3\%)\\
   \midrule
    Does not exist & Contains unpinned dep.  & 277 (24.2\%)\\
    Does not exist & Contains pinned dep. & 135 (11.8\%)\\
  \bottomrule

\end{tabular}
\end{table}

As can be seen in Table~\ref{tab:smells}, the URL dependency smell is less common but still appears in over 10\% of the projects. This is followed by the restrictive constraint smell which appears in more than 9\% of the projects. The Permissive constraint is perhaps the least common smell, appearing in less than 4\% of the projects. This makes sense since this smell is almost never justified, unless we have complete control over the dependent package. 
Our results show that developers are much more likely to err on the side of caution when it comes to breaking changes, since restrictive constraints and pinned dependencies are substantially more common than permissive constraints. 

Unused and missing dependencies are surprisingly common, appearing in 79\% and 63\% of the projects, respectively. 
While the high incidence of unused dependencies indicates an overall lack of attention to dependency maintenance and can bloat the installation of some applications, our analysis show that the large majority of the projects require dependencies that are not declared in the package.json (missing dependencies), which can cause bugs and crashes. 
Overall, these results indicate JavaScript developers seem to struggle to maintain a healthy package.json file. 

\rev{We studied the relationship between project characteristics such as age, number of dependencies, number of commits, and the number of maintainers on the occurrence of each dependency smell. Based on the mentioned characteristics, we found significant differences between smell-infected and smell-free projects using the Man-Whitney-U test for p$<$0.05. However, analyzing the effect size using Cliff's delta reveals the differences to be small or negligible for all characteristics except the number of dependencies (large effect size). This is not surprising as increasing the number of dependencies increases the likelihood of a project being infected by a smell.}\\
So far, we have investigated whether projects contain dependency smells \rev{or whether smell occurrence is affected by project characteristics,} but not the extent in which they occur within the projects. 
For instance, pinning a single dependency may have much less impact on a project maintenance than pinning half of its project dependencies. 
To put the extent of smell occurrences per project into perspective, we show in Figure~\ref{fig:dist-smells} the distribution of \textit{dependency smell percentage} on the projects with at least one instance of that smell. 
The dependency smell percentage normalizes the number of dependency smells with respect to the total number of the project dependencies, as projects with large number of dependencies will have a tendency to have more smells.
In the case of Missing Dependency, we perform a different normalization, as these represent dependencies that are not specified by the project.
We calculate the dependency smell percentage by dividing the number of identified missing dependencies over the number of dependencies in package.json + the number of missing dependencies.
Note that we did not include the no package-lock smell, as calculating the  percentage for this particular smell makes no practical sense - a project either has the smell (100\%) or not (0\%). \\
\indent{All of the violin plots have a relatively non-uniform distribution, with the majority of smell distributions having a median of under 17\%. In other words, smells usually infect a minority of a projects dependencies.}
In particular, projects tend to have a relatively small percentage of URL and Missing dependency smells, while Pinned and Unused dependencies occur in higher proportions in our dataset. 
Observing the pinned dependency plot near the 100\% mark shows that some projects (7.5\% of all projects) may decide to use pinning as a blanket dependency strategy regardless of the specifics of each dependency. On the other hand, the URL dependency smell is often used on specific cases. We further explore the reasons why these occur in Section \ref{section:rqiii}.

\conclusion{\rev{The results reveal that 80\% of the projects are infected with two or more distinct smells. However, not all smells occur at a large degree. Four out of seven appear in less than 30\% of projects and the majority of smells infect a minority of a project's dependencies. }}

\subsection{\rqii}

\noindent\textbf{Motivation:}
After realizing the prevalence of dependency smells in JavaScript projects, we want to understand how developers perceive these smells and their negative consequences. Do the developers agree that these smells are harmful? Do different smells have different levels of impact? This will help us understand if the prevalence of these smells are in fact a major concern.

\noindent\textbf{Approach:}
In order to examine the negative consequences of dependency smells, we crafted a survey and asked respondents to rate their agreement/disagreement to a number of statements about each smell. 
We craft the statements by grouping the responses from the first survey from Section \ref{sec:Dependency smells} into encompassing statements about each smell. 
We also included statements about using SemVer, to assess how developers see the use of this standard in practice.  
 
To that aim, we create a questionnaire containing all statements in \Cref{tab:initial_sur}, and ask practitioners to rate their agreement on a likert-scale of 1 to 5 (strongly disagree, disagree, neither, agree, strongly agree). \rev{Note that some statements might seem positive/negative, but considering the level of disagreement from developers may reveal the opposite. For example, the positive statement that restrictive constraints ``allow for all important security and bug fixes" has been met with disagreement, meaning the inverse is true.}
We were careful not to bias the participants in being negative towards the dependency smells.
To do so, first, we refer to each of the smells as dependency approaches and include the SemVer compliant approach in the questionnaire. 
Second, to avoid confusing the participants, we did not use the labels used throughout this paper (e.g., Restrictive constraint, URL Dependency) and instead focus in describing the smells as an ``approach" for managing dependencies.
For instance, instead of writing ``Restrictive Constraint'' we specified ``Allowing \npm to only install the latest patch updates''.
Instead of "URL Dependency" we use ``Specifying a direct URL to fetch the dependency'', and so on.
The sole exception are the dependency smells Unused Dependency and Missing Dependency, which are clear dependency issues and we framed them as such in the questionnaire.
The participants could also freely input comments on a text box for each dependency smell, to clarify and justify their agreement or further address aspects not mentioned in our questionnaire.

Similar to Section~\ref{sec:Dependency smells}, we ask the advantages and disadvantages
of \textit{using} package.lock in the project, given this
more intuitive and easier to assess than the consequences
of not using an approach. We then use the pros of using a
package.lock as the cons of not having package.lock in the
project, and vice versa.

It is worth noting that we did not include statements that relates to how SemVer or \npm operates, since our goal was to evaluate qualitative statements rather than quiz the respondents on how SemVer or \npm works. For example, a common issue associated with pinned dependencies is that developers cannot automatically receive new updates. However, this is a fact and is not up to debate. Even though it exists in our smell definition, it does not appear in the survey. On the other hand, the security problems associated with pinning dependencies is a qualitative characteristic which depends on developers' experience.

To recruit our survey participants, we use a snowball sampling method~\cite{goodman1961snowball} where we posted this survey on Twitter and the Node.JS community on Spectrum. 
We received a total of 41 replies.
As \Cref{tab:survey-background} shows, the set of respondents is composed primarily by industry practitioners (75.6\%) and academic researchers 19.5\%).
The majority of the participants have more than a year of Node.JS development experience (83.3\%), use \npm on most or all of their projects (92.7\%) and are completely familiar with \npm (80.5\%). We present the background and the experience of respondents with Node.JS, \npm and SemVer in \Cref{tab:survey-background}.

\begin{table}
	\centering
	\caption{Background of participants in the survey.}
	\label{tab:survey-background}
	\begin{tabular}{p{1in}p{1.2in}r}
	
	\toprule
	\textbf{Dimension}	& \textbf{Experience} &  \textbf{\%}\\
	
	\midrule
	\multirow{3}{*}{Background}		
	& Industry Practitioner				&		75.6\%	\\
	& Academic Researcher			&		19.5\%	\\
	& Student								&	4.9\%		\\

	\midrule 
	\multirow{4}{*}{Node.JS}		
	&	$>$ 5 years						&			24.4\%	\\
	&	4-5 years						&			24.4\%	\\
	&	1-3 years						&			41.5\% 	\\
	&	$<$ 1 year						&			9.8\%	\\

	\midrule
   \multirow{4}{*}{\npm use}			
	&	Always							&		 48.8\% \\
	&	Most projects				&		43.9\% \\
	&	Sometimes						&		 4.9\% \\
	&	Rarely								&		 2.4\%\\
	\midrule
	\multirow{3}{*}{SemVer}
	& Completely familiar 				&		 80.5\%	\\
	& Somewhat familiar 				&		 12.2\%	\\
	& Not familiar 							&			7.3\%\\
	
	\midrule
		\multicolumn{2}{l}{\textbf{Total participants}}  & \textbf{41} \\
	
	\bottomrule
\end{tabular}
\color{black}

\end{table}

\begin{table*}[tbh]
	\centering
	\caption{Survey results for quantifying smell characteristics. Each statement is either a rationale ($+$) for using the smell as a valid approach or a downside ($-$) of the dependency smell. The last column presents the aggregated total for each statement from 1 to 5 (strong disagreement to strong agreement)}
	\label{tab:initial_sur}
	\includegraphics[width=\linewidth]{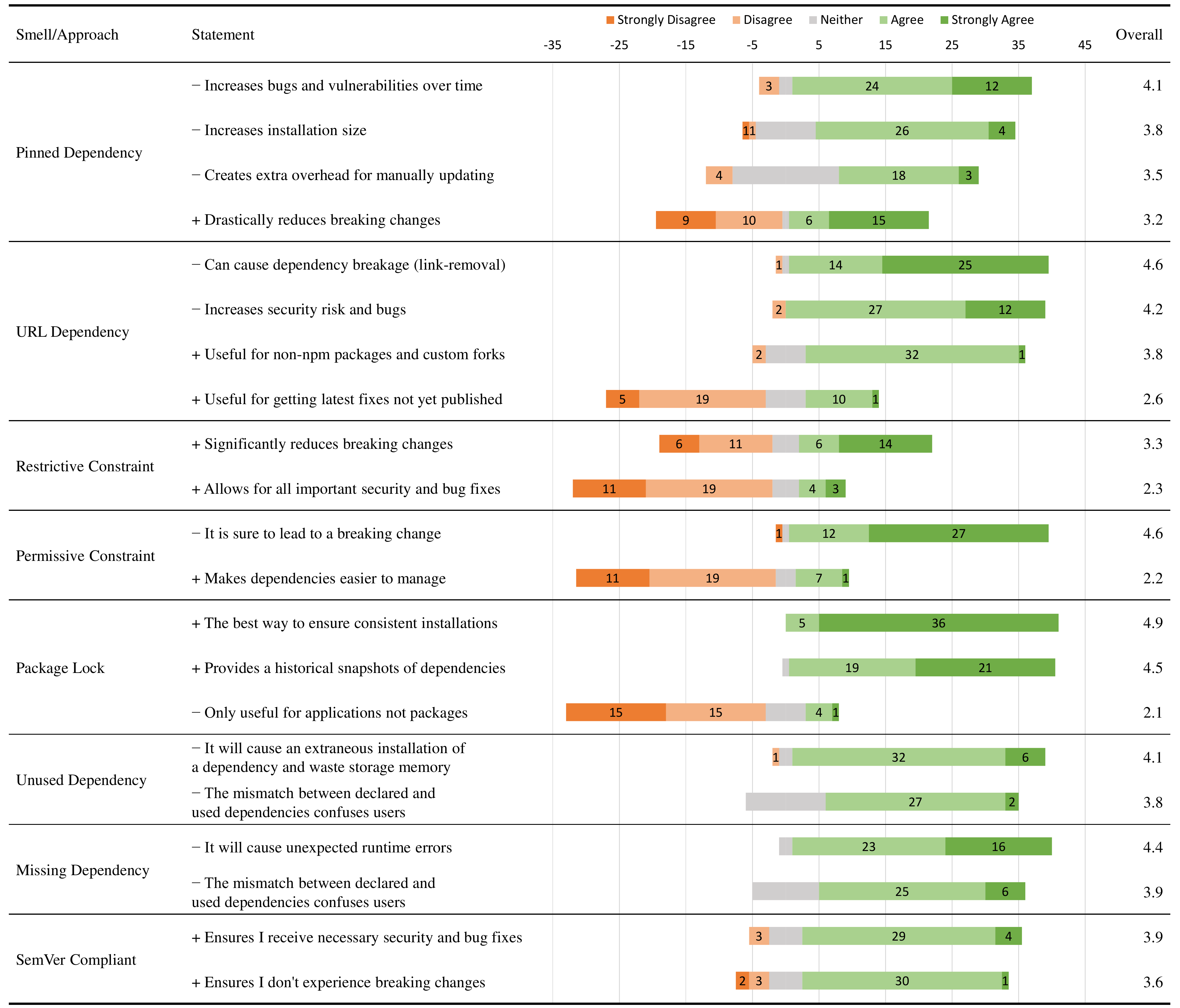}
\end{table*}

\noindent\textbf{Results:}
We present the results of our survey in \Cref{tab:initial_sur}, showing per dependency smell and statement, the distribution of the agreement levels as a Divergent Stacked Bar chart.  
We center-align the bars on ``Neither''. Hence, a statement with mostly green bars tending to the right shows a more frequent agreement within the 41 participants. 
Conversely, statements with bars tending to the left (orange) convey a more frequent disagreement to the statement by the respondents. 
Note that each statement is either a rationale for using the dependency smell/approach (marked with `$+$') or a disadvantage of using the smell/approach as a valid approach (marked with `$-$').

Overall, the results shown on the third column of \Cref{tab:initial_sur} indicates that developers substantially agree with 15 out of the 21 statements.
Developers mostly agree with all statements related to the benefits of using SemVer, and the problems that may be caused by Unused and Missing Dependencies. 
Aside from this, we see a more common agreement on the downsides of using dependency smells.
For instance, the three highly agreed upon statements related to Pinned Dependency are all related to the downsides of pinning dependency: increases bugs and vulnerabilities over time, increases installation size and creates extra overhead for maintainers.
We observe a similar picture in the highly agreed upon statement for Permissive Constraint and URL Dependency, indicating that {developers tend to agree with the downsides of using the dependency smells}.

All the six statements where {developers more frequently disagree are related to the rationale of using a dependency smell as a valid approach}.
For instance, most developers either disagree or strongly disagree that using a Permissive Constraint makes dependencies easier to manage.
When justifying their position, developers pointed out that: \textit{``It is only easier to manage [the dependencies] if nothing breaks, which it will''} and said that: \textit{``Solves some vulnerabilities but causes others.''}
Another vastly disagreed upon statement, is related to Restrictive Constraint: Allows for all important security and bug fixes.
Developers pointed out that: \textit{``[...] maintainers bundle many fixes into non-patch releases"} and said that: \textit{``(one) should never ONLY rely on patch releases for fixes."}
Another highly contested view is related to the cons of using package.lock: Only useful for applications not packages.
As we mentioned before, this statement actually points out a disadvantage of not using package.lock in projects.
Most practitioners disagree, pointing out that: \textit{``Even for packages, it decreases test flake in CI, causing less maintainer stress"}, showing that they believe there is not a good reason not to include a package.lock on JavaScript projects.

Overall, developers are even more conservative when it comes to rationalizing the usages of dependency smells than we anticipated.
While all the downsides of using dependency smells as an approach were highly agreed upon by practitioners, they have disagreed with statements that attempt to rationalize dependency smells as valid approaches in particular cases.

\conclusion{Not only do practitioners confirm that dependency smells can be harmful to the maintenance and security of software projects, but practitioners are even critical towards rationales for turning to these smells in specific circumstances.}

\subsection{\rqiii}
\label{section:rqiii}

\noindent\textbf{Motivation:}
Observing the prevalence and the impact of dependency smells leads us to ask the question: why are these smells introduced? We were keen to identify reasons on why developers have introduced a certain dependency smell and whether the introduction of a smell was a conscious effort by the developer or if it was merely a by-product of other development or maintenance tasks. 

\noindent\textbf{Approach:}
For this analysis, we reach out to 100 developers that have introduced a dependency smell in the studied projects in our dataset. 
We only consider the cases where a developer changed the dependency constraint from SemVer to a dependency smell. In such cases, we expect developers to be aware of semantic versioning and are more likely to remember what motivated them to use an alternative approach. We also prioritize more recent changes with the hope that the changes are still relatively fresh in their minds (although this resulted in an older change from 2015 for the permissive constraint smell to also be included, since this smell has less cases overall). We ignore "no-reply" emails.

In total, we sent out 100 emails that reached a recipient, 25 for each of the 4 dependency smells: Pinned dependency, URL dependency, Restrictive constraint, and Permissive constraint.
We omit Missing and Unused dependencies from this analysis, as these are clearly a symptom of dependency mismanagement and there are no valid rationales for introducing them.
We also omit "No Package-Lock" from this analysis as this is a project decision not easily attributed to a single developer we could contact to get information. It is also not a decision that regularly changes due to dependency modifications and is often related to the development culture of individual projects.
Our email provided details of the change such as the dependency name, dependency constraint, and the commit message, along with GitHub links to the commit that caused the change. We asked developers "why they switched to this approach instead of using SemVer like before". 
Developers were suggested to justify their changes by simply replying to the email.

We received a total of 28 responses (i.e., response rate of 28\%) where developers explained their reasoning for the change. The achieved response rate is significantly larger than the typical 5\% rate found in questionnaire-based software engineering surveys \cite{singer2008software}. This is because developers were targeted based on their dependency actions and we also tried to provide all the relevant information in the email while keeping it concise and to the point.
The responses were quite detailed in a free-text format, providing not only the reasoning about the dependency change in question, but also rationales on why the developers might resort to dependency smells, along with cultural beliefs about semantic versioning and the centralized nature of \npm.

To analyze the free-text responses provided by the developers, the first two authors have independently assigned a theme to each reason presented in the responses, using an open-coding approach~\cite{Fincher_2005OpenCardSort}.
The two authors then discussed the themes encountered and merged them into 10 categories of reasons of why the dependency smells are introduced. 
In the two cases where the first two authors did not fully agree on the classification, the third author was consulted to provide a tie-breaker.

\noindent\textbf{Results:}
The 10 reasons on why dependency smells are introduced are presented in \Cref{tab:dev_sur}. 
Note that the total number of responses in the table is more than 28, because many developers gave multiple reasons for introducing the dependency smells.
We found at most four different reasons per dependency smell, while the most common reason on each smell was found in at least three responses. 
In the following, we explain each reason along with example responses from developers (Tagged P1 through P28):

\begin{table}[tbh]
	\centering
	\caption{Developer reasons on why dependency smells are introduced.}
	\label{tab:dev_sur}
	
\begin{tabular}{l l r}
		\toprule
		\textbf{Smell} & \textbf{Reason} & \textbf{\# Responses}\\
		\midrule
		\multirow{4}{1.5cm}{Pinned Dependency} 
		& Breaking changes & 5 (71\%) \\
		 & Mistrust of SemVer compliance & 3 (43\%) \\
		 & Transitive dependency compatibility & 1 (14\%) \\
		 & Unaware of SemVer & 1 (14\%) \\
   		\midrule
   		
		\multirow{4}{1.5cm}{URL Dependency}
		& Fix not on \npm & 7 (70\%) \\
		 & Experimenting with features & 4 (40\%)\\
		 & Philosophical issue & 3 (30\%)\\
		 & Maintainer owns dependency & 2 (20\%)
		\\
		\midrule
		
		\multirow{3}{1.5cm}{Restrictive Constraint} 
   		& Breaking changes & 5 (83\%)\\
   		& Mistrust of SemVer compliance & 2 (33\%)\\
   		 & Transitive dependency compatibility & 1 (17\%)\\
   		\midrule
   		
   		\multirow{3}{1.5cm}{Permissive Constraint} 
   		 & Breaking changes not likely & 3 (60\%)\\
   		 & Unaware of SemVer & 1 (20\%)\\
   		 & Reducing installation size & 1 (20\%)\\
		\bottomrule
	\end{tabular}
\end{table}

\newcommand{\rqquote}[1]{\textit{#1}}
\noindent\textbf{R1-Breaking changes:}
The most cited reason for pinning dependencies and also for using restrictive constraints is breaking changes. \rev{Breaking changes are changes in a new release of a package that is incompatible with the previous API, causing a breakage in others that depend on the package.} Developers tend to adopt a more cautious approach when a dependency breaks the API even though it was a minor or patch release. For example, P12 said: \rqquote{``newer versions of "esm" broke GL JS tests in very obscure ways. I spent a day debugging and couldn't find a root cause for''}. It is also interesting to see that new features, while not causing API breakage, can still cause problems, as stated by P13: \rqquote{``a new feature introduced in a minor release might be backwards compat in terms of the API, but not so much in terms of the user experience or expectations - e.g. a component might suddenly show some new option or expose some new behavior that we have yet to evaluate or have approved by our clients / end users''}. Developers are in fact opting to pin dependencies as a direct response to breaking changes or unexpected updates.

\noindent\textbf{R2-Mistrust of SemVer compliance:}
Some developers have not experienced breaking changes for the specific dependency, but past experiences with other dependencies have made them less trusting in SemVer and more conservative in their dependency approach. For example, P27 mentions:  \rqquote{``I think it is a result of several incidents with packages and my mistrust of [the] js community''}. In fact, not trusting dependency maintainers in \npm to properly follow SemVer is the second highly cited reason for pinning a dependency or using a restrictive constraint, as cited by P16: \rqquote{``The unfortunate reality is that not every maintainer correctly adheres to SerVer when publishing new versions of their own dependencies"}. Past experiences leave a bitter taste and it may take a while for some developers to opt for SemVer even after maintainers properly abide by it.

\noindent\textbf{R3-Transitive dependency compatibility:}
Sometimes transitive dependencies can cause incompatibility issues. Such cases arise when a transitive dependency requires only specific versions of a dependency to be installed, whereas \npm will install multiple versions of a dependency if they are required by dependency packages. For example, P22 said: \rqquote{``So to ensure that a transitive dependency issue do not occur I pinned the react-scripts version"}. Although developers are somewhat aware that this issue could be remedied by using peerDependencies \rev{(plugins that express compatibility with a specific host package but do not use it directly through the ``require" statement \cite{npmpkg2019})}, as stated by P11: \rqquote{``it’s probably better to express the dependencies within the hierarchy as loose  peer-dependencies"}.

\noindent\textbf{R4-Unaware of SemVer:} While not common, some developers mentioned their lack of familiarity with SemVer as the reason for pinning a dependency or using a permissive approach, as cited by P18: \rqquote{``I really had no idea what the hell I was doing back in 2015, and had probably not yet even heard of semver"}. Some dependencies can still suffer from old practices in which SemVer was not a well-known standard.

\noindent\textbf{R5-Fix not on \npm:} The most common reason for using a direct URL to fetch a dependency is \rev{the delay or reluctance of the maintainer in publishing a needed fix on \npm}. For example, P9 stated: \rqquote{``There is a bug in the main npm module and there's a branch made to fix it but it hasn't been merged."}. In some cases, the developers even issued a fix to the repository but the maintainer did not respond, as mentioned by P2: \rqquote{``I submitted a pull request to fix the vulnerability, but they never bothered merging it, and their system eventually closed the PR because it sat for too long"}. If the maintainers would work with users to publish a consistent and timely stream of patch releases with the newest fixes, there would be less incentive to circumvent \npm to fetch dependencies.

\noindent\textbf{R6-Experimenting with features:} Another common reason for switching to a direct URL is to experiment with new features not yet ready for release, as mentioned by P9: \rqquote{``I'm experimenting, and not sure if certain things are stable yet"}. This experimentation is also useful if the developer wants to prepare for an upcoming transition to a new release. For example, P8 said: \rqquote{``At the time I did that because I wanted to test against a specific branch"}. Some developers are fully aware of the unstable nature of using unpublished dependency versions and view it as a temporary experimentation.

\noindent\textbf{R7-Philosophical issue:} Interestingly, some developers have an inherent and somewhat philosophical issue with the idea of a centralized dependency registry. For example, P8 stated: \rqquote{``I don't think it is beneficial to have a standard centralised registry for dependencies in any language. That's what \npm has become to node and it is critical and prejudicial, as npm is a private company and can choose to do whatever it wants"}. In such cases, they believe using absolute URLs for JavaScript dependencies is a major step towards decentralization, as pointed out by P14: \rqquote{``since JavaScript is an interpreted language I have a hunch that it's actually not a good thing to use a centralized registry"}. The philosophical standpoint against a centralized dependency model is an interesting and unexpected reason why some developers may never fully depend on \npm, but such discussions often ignore the disadvantages of using URLs, such as link breakage and typo-squatting attacks.

\noindent\textbf{R8-Maintainer owns dependency:} We also observed that using a URL to use a development branch of a dependency managed by you can be useful for integration tests. For example, P26 cited: \rqquote{``When we are developing something in \rev{`[anonymised]'} or in any other dependency managed by us which are npm dependencies, we used to point to the Github development branch of the dependency in order to a more or less detailed validation of the changes"}. It is also useful if it is crucial to have a personal fork in order to have full control over the dependency. For example, P28 mentioned: \rqquote{``We switched to our own forked version of the repo. This is to ensure that it works properly, that no vulnerabilities are introduced (since our app deals with people's money)"}. The main advantages and disadvantages of using URLs relates to "trust", but some developers  trust a custom URL, or even trust their forks more than the official package due to increased control.

\noindent\textbf{R9-Breaking changes not likely:} Contrary to reasons 1 and 2, some developers have not experienced breaking changes. This is the most cited reason for using permissive constraints, as stated by P7: \rqquote{``I might have done it differently if it was likely that we'd break people with this, but it just didn't seem likely."}. However, developers mention how this only works in specific circumstances when there is a high familiarity with the dependency, as mentioned by P25: \rqquote{``while it works perfectly for our use case, this situation is not ideal from a semantic versioning perspective"}. Just as negative experience can shape one to rarely trust the compatibility promises of SemVer, those who have never experienced breaking changes may not be worried about their consequences.

\noindent\textbf{R10-Reducing installation size:} Another reason for using permissive constraints is so that users of your package can reuse dependency versions without \npm installing multiple versions of the same package. For example, P7 mentioned: \rqquote{``allowing more permissive use may allow users to reduce the total number of packages that they need to ship in their bundles"}. In these cases, if you already have a particular version of package A in your dependencies, and package B  in your dependency tree relies on package A, it is more likely that package B does not need to install a separate version of package A. While overly permissive approach can create many issues, being overly restrictive could force clients into larger installations as \npm cannot fully utilize reusability and needs to install multiple versions of the same package.

\conclusion{The introduction of dependency smells is generally the result of developers reacting to dependency misbehaviour and the shortcomings of the \npm ecosystem. Nonetheless, some developers are unaware of SemVer and some are against a centralized dependency model altogether.}

\color{black}
\section{Dependency Smell Evolution}
\label{sec:Discussion}

\begin{figure*}
  \begin{subfigure}[b]{0.25\textwidth}
    \includegraphics[width=\textwidth]{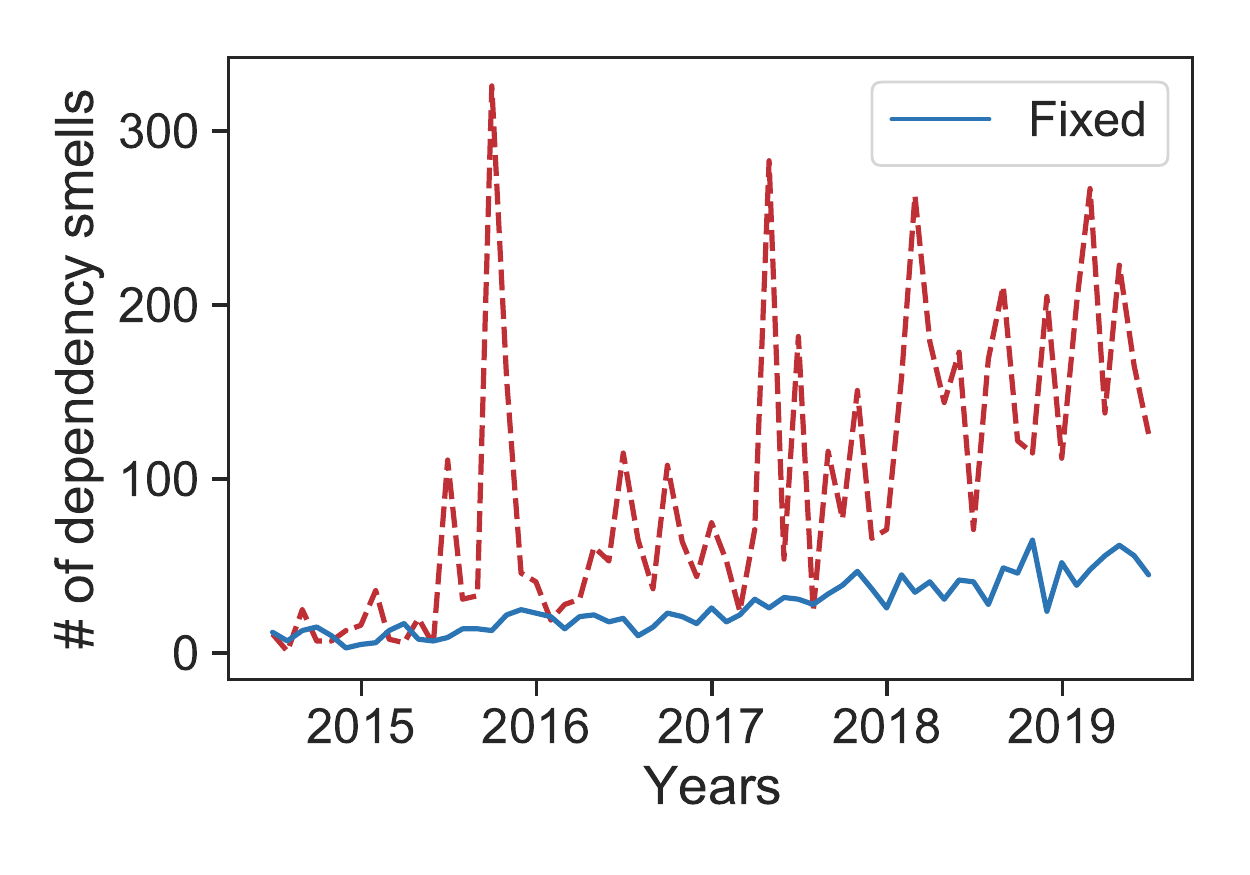}
    \caption{\small Pinned Dependency}
    \label{pinned-introduced-fixed}
  \end{subfigure}
  \hspace{-1mm}%
  \begin{subfigure}[b]{0.25\textwidth}
    \includegraphics[width=\textwidth]{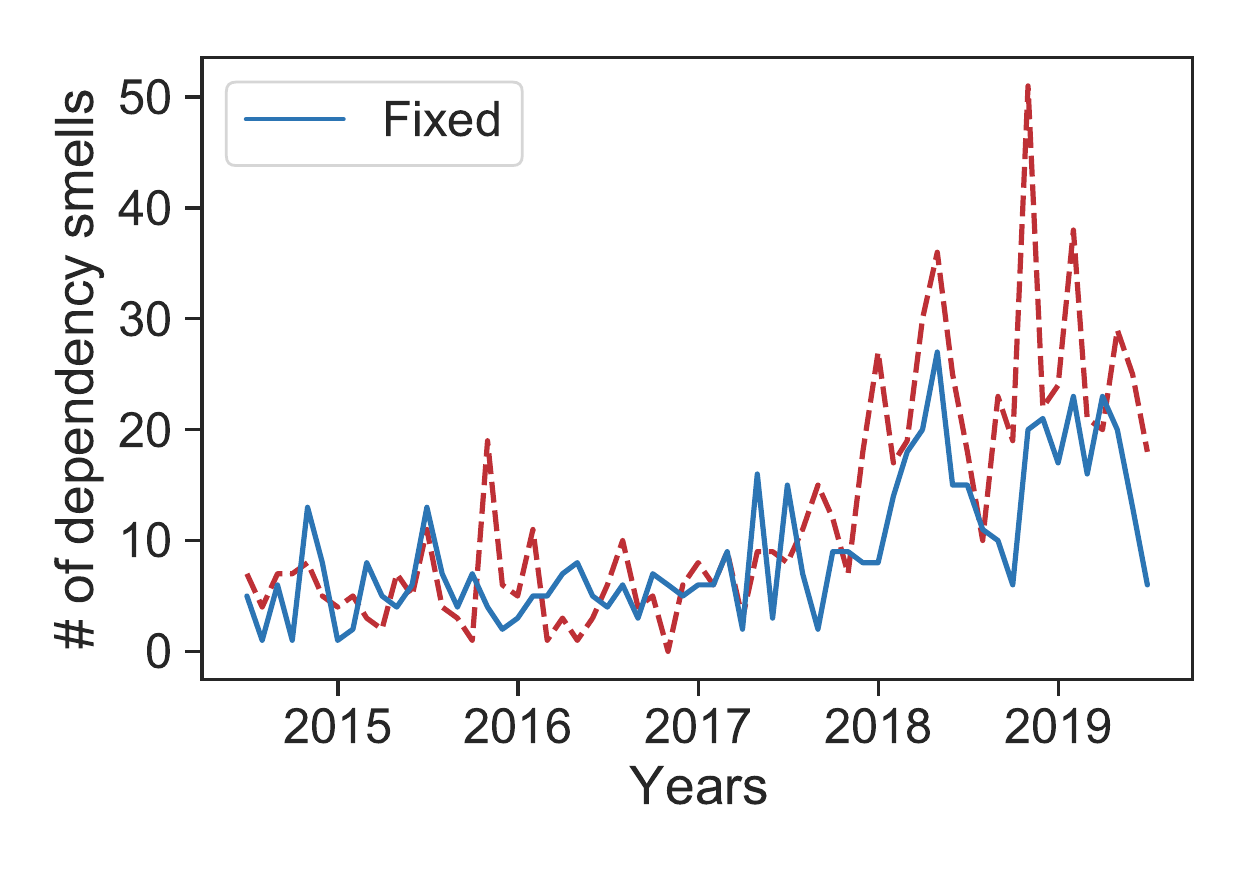}
    \caption{\small URL Dependency}
    \label{url-introduced-fixed}
  \end{subfigure}
\hspace{-1mm}%
  \begin{subfigure}[b]{0.25\textwidth}
    \includegraphics[width=\textwidth]{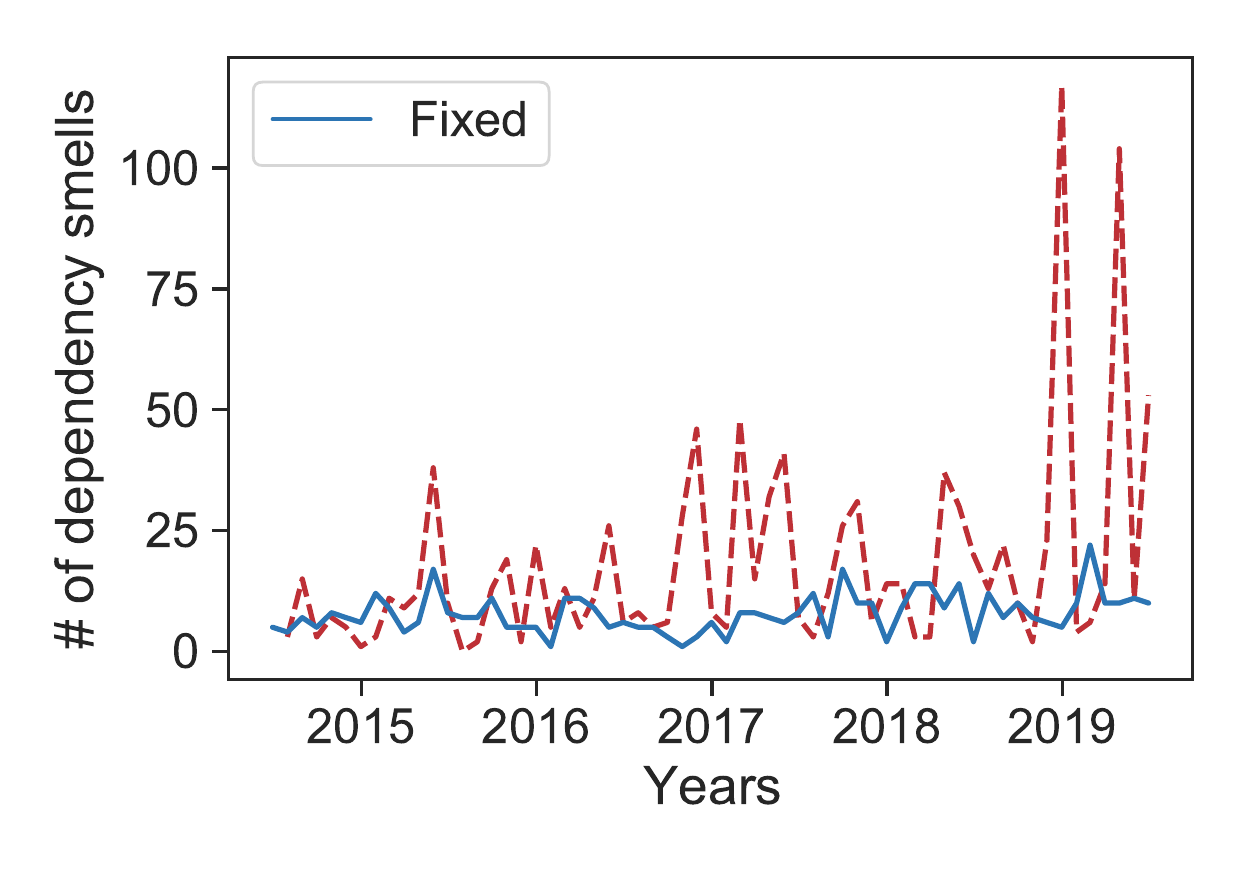}
    \caption{\small Restrictive Constraint}
    \label{restrictive-introduced-fixed}
  \end{subfigure}
  \hspace{-1mm}%
  \begin{subfigure}[b]{0.25\textwidth}
    \includegraphics[width=\textwidth]{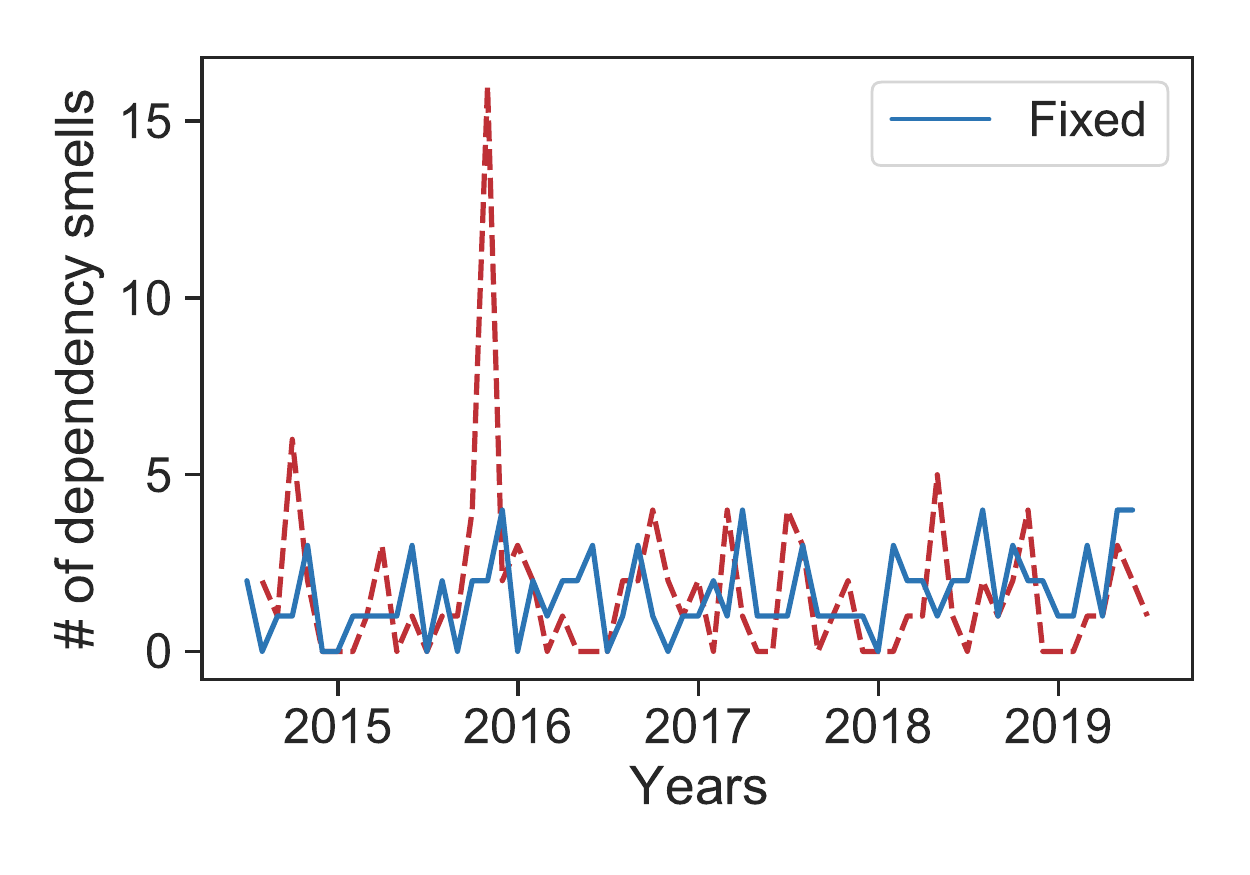}
    \caption{\small Permissive Constraint}
    \label{wildcard-introduced-fixed}
  \end{subfigure}
\caption{Introduced and fixed dependency smells over time.}
\label{fig:smells-introduced-fixed}
\end{figure*}

\begin{figure*}
	\begin{subfigure}[b]{0.25\textwidth}
		\includegraphics[width=\textwidth]{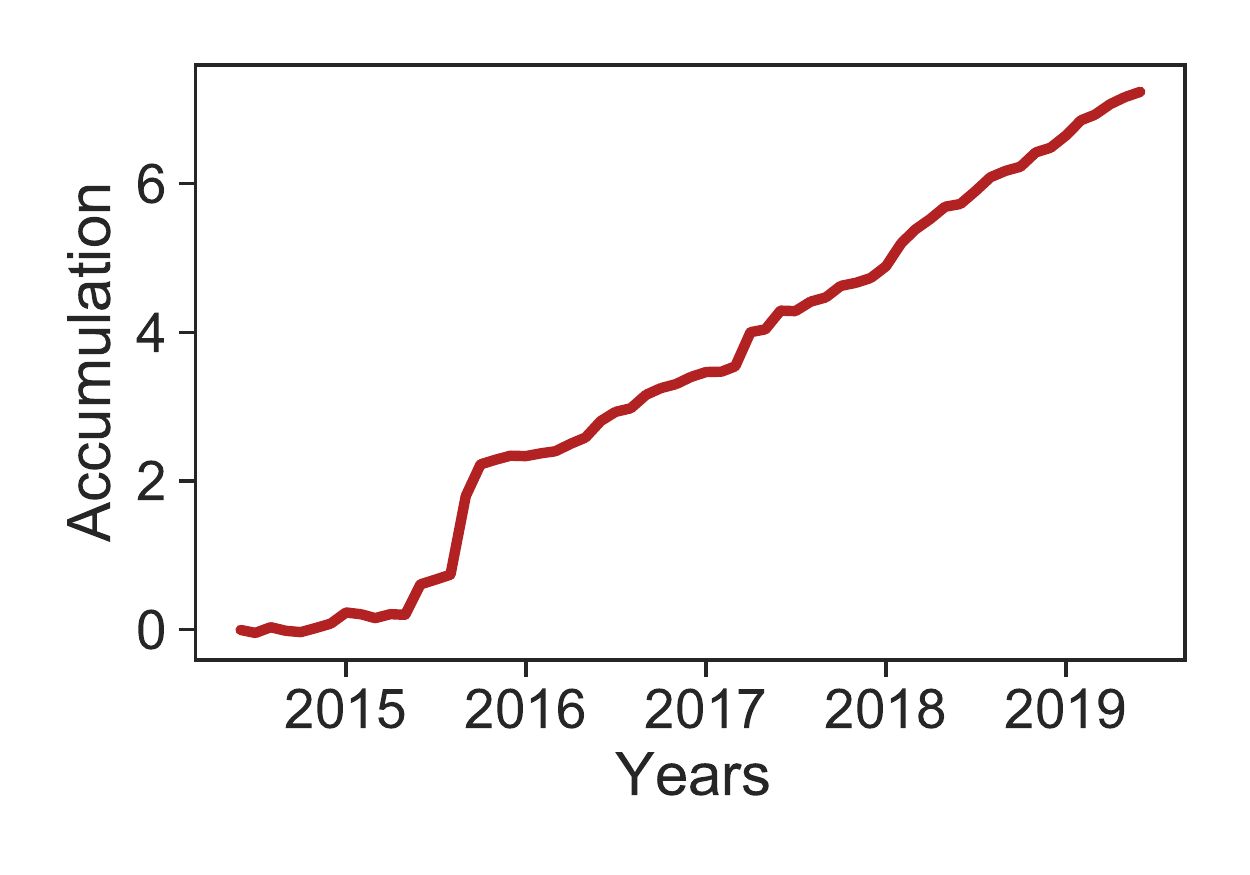}
		\caption{\small Pinned Dependency}
		\label{pinned-accumulation}
	\end{subfigure}
	  \hspace{-1mm}%
	\begin{subfigure}[b]{0.25\textwidth}
		\includegraphics[width=\textwidth]{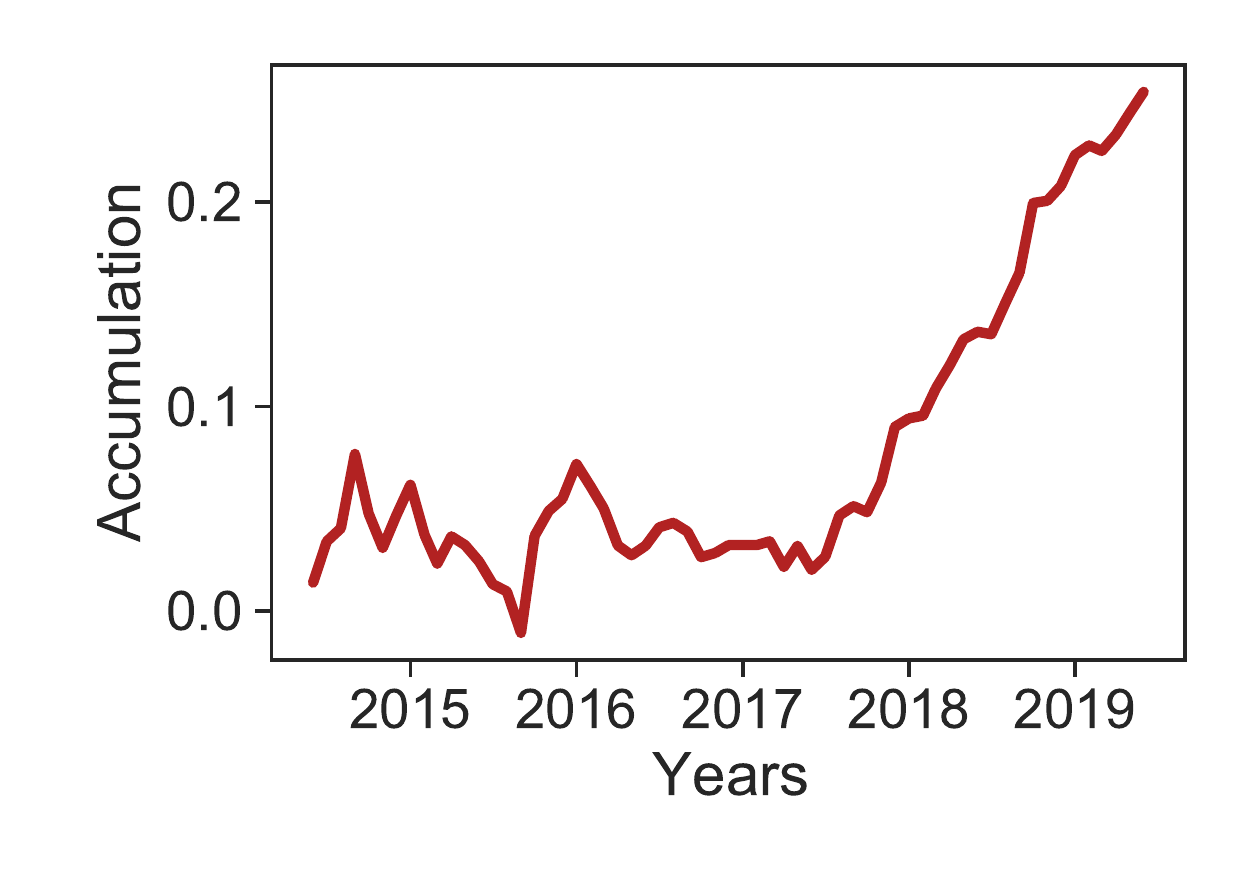}
		\caption{\small URL Dependency}
		\label{url-accumulation}
	\end{subfigure}
	  \hspace{-1mm}%
	\begin{subfigure}[b]{0.25\textwidth}
		\includegraphics[width=\textwidth]{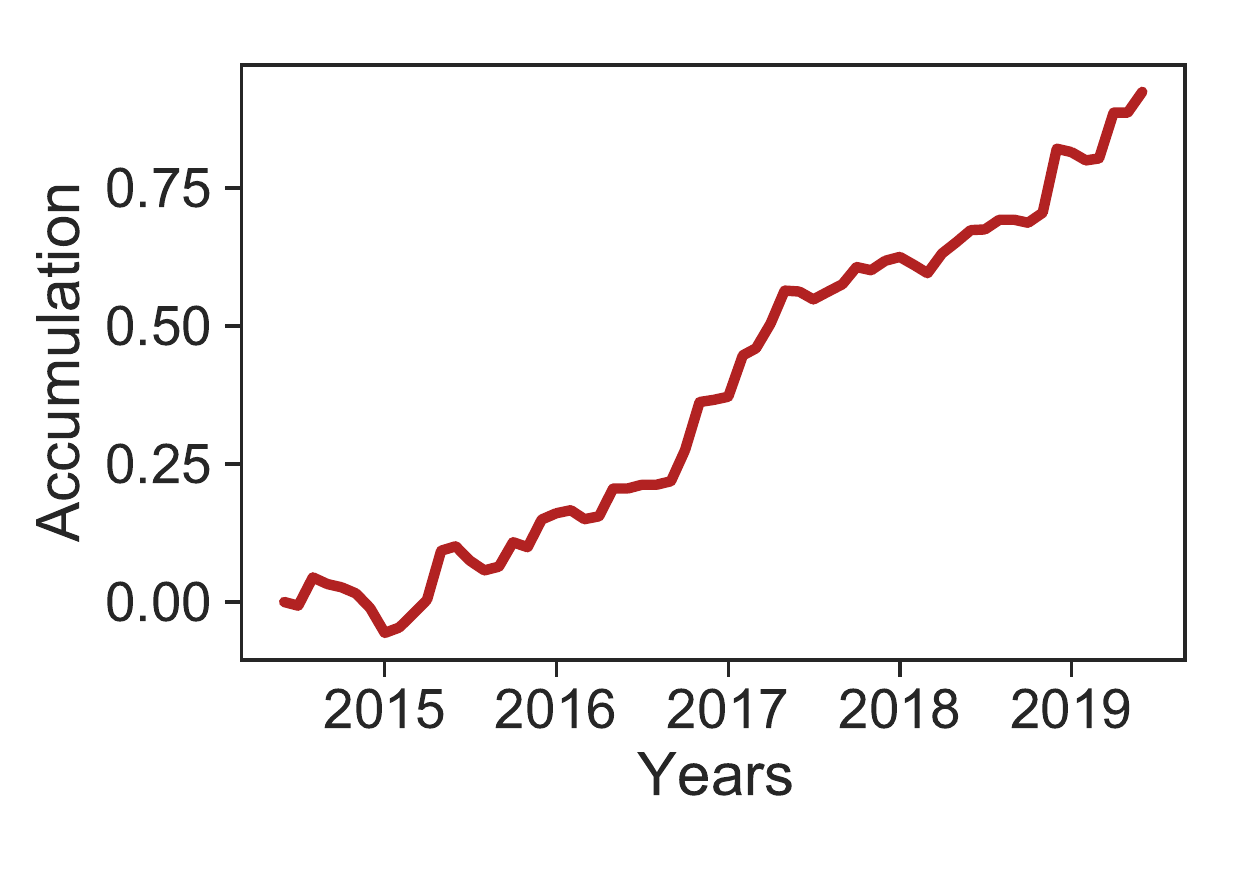}
		\caption{\small Restrictive Constraint}
		\label{restrictive-accumulation}
	\end{subfigure}
	  \hspace{-1mm}%
	\begin{subfigure}[b]{0.25\textwidth}
		\includegraphics[width=\textwidth]{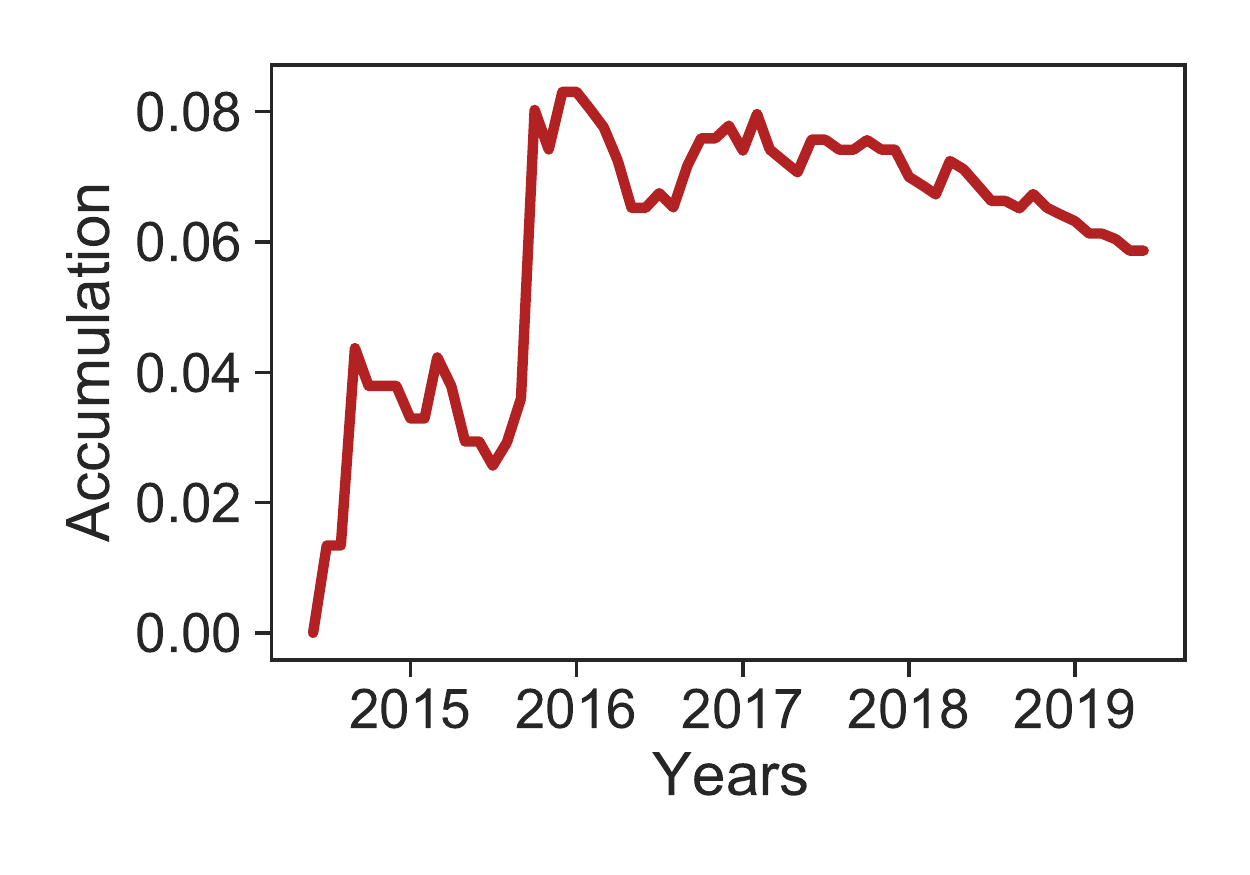}
		\caption{\small Permissive Constraint}
		\label{wildcard-accumulation}
	\end{subfigure}
	\caption{Accumulation of dependency smells over time.}
	\label{fig:accumulation}
\end{figure*}

Now that we know dependency smells are prevalent, we want to investigate smell-inducing and smell-fixing commits throughout history to better understand their evolution. By analyzing the smells that were introduced or fixed in the past five years, we can better understand when these smells started to accumulate and where the current trend is moving towards.

Since we have the contextual diff for all commits parsed and stored into our database, we can identify when and how dependency constraints were changed throughout the project. We specifically focused only on dependencies that persist between two commits but their constraint was changed such that a dependency smell was introduced or fixed. This prevents noisy data due to fixes that are merely a result of dependency removal or migration. Since the analyses for this research question rely on the dependency constraints inside the package.json file, it can only be conducted on the first four dependency smells (pinned dependency, URL dependency, restrictive constraint, and permissive constraint). We have analyzed the changes over a 5 year period to provide a more comprehensive view of how the smells evolve through time.

\begin{table}
	\centering
	\caption{Number of introduced and fixed instances per dependency smell.}
	\label{tab:instances}
		\begin{tabular}{clrr}
		\toprule
		\multirow{2}{*}{\textbf{\#}}					& 	\multirow{2}{*}{\textbf{Smell}}	& \multicolumn{2}{c}{\textbf{~~\underline{Smell Instances}}} \\
		 &  & \textbf{Introduced~} & \textbf{~Fixed}\\
		\midrule
		S1 & Pinned dependency & 5,814 & 1,640\\
		S2 & URL dependency & 750 & 556\\
		S3 & Restrictive constraint & 1,068 & 479\\
		S4 & Permissive constraint & 99 & 96\\    
		\bottomrule
	\end{tabular}
\end{table}

Table~\ref{tab:instances} presents the total number of instances for each of the four smells that were introduced or fixed during a five-year period from mid 2014 to mid 2019. The line plots in Figure~\ref{fig:smells-introduced-fixed} show how the introduction and fix of dependency smells happen throughout project evolution. The fixed lines clearly show that developers are somewhat aware of these smells and they regularly address them.

In order to investigate whether the regular fixes are enough to alleviate the regular introduction of dependency smells, we have plotted the accumulation of these dependency smells in Figure~\ref{fig:accumulation}. \rev{This accumulation is normalized by the number of projects included in the dataset per month (since projects are added at different times).} As can be seen, the pinned dependency, URL dependency, and restrictive constraint smells have an increasing trend. This shows that despite consistent dependency refactoring throughout the history of these projects, the number of newly added dependency smells are still higher.

The accumulation graph for the permissive constraint smell is noticeably different than the rest, both in its spiky form and its recently downwards trend. The non-gradual shape of the diagram might be due to the smaller number of instances for this smell (Table~\ref{tab:instances}). The permissive constraint smell subjects projects to a higher risk of breaking changes and the recently decreasing trend implies that developers generally agree that this smell is not justified, or the drawbacks far outweigh any benefits. It may be for this reason that another ecosystem like Cargo has banned the use of permissive wildcards in 2016 \cite{cargowild2018}. \rev{It may also be due to the severity of this smell. According to  \Cref{tab:initial_sur}, Permissive Constraints have the highest likelihood of introducing breaking changes. Another reason for this downward trend could be the introduction of the semver-compliant caret (\textsuperscript{$\wedge$}) symbol in \npm and its use as the default in February 2014 \cite{decan2019, nodesource}. This gives developers more freedom by allowing minor and patch updates (as opposed to only patch updates) and discourages the use of more permissive constraints..}

Dependency smells are addressed and fixed through time, but most dependency smells are introduced more frequently than they are fixed, causing an accumulation of smells over time. The permissive constraint smell is the only case where we observed a decrease over time.

\section{Generalizability to other ecosystems}
\label{sec:Discussion}
While our study focuses on \npm, many of these smells can apply to other ecosystems with varying degrees. In fact, any package manager that allows developers to restrict dependencies to a particular version or version range is susceptible to pinned and restrictive dependency smells. It will also be susceptible to permissive and URL dependency smells if the package managers allow developers to always use the latest version of a package or fetch a dependency directly from a URL. Missing dependencies (and to a lesser extent, Unused dependencies) are mainly a problem for interpreted languages such as JavaScript and Python. However, compiled languages such as C++ (and even languages like Java) can catch these issues at compile time.
For example, the PyPI ecosystem for Python allows developers to specify dependencies (requirements specifiers) and constraints (version specifiers) in a setup.py or requirements.txt file. The Python dependency specification \cite{pipsemver} allows for a pinned dependency using the version matching clause (==). PyPI also uses the compatible release clause ($\sim=$) and the ordered comparison clause ($>$= and $<$=) which allows for the restrictive constraint and permissive constraint smells. Python projects can also suffer from unused and missing dependencies since one can specify dependencies regardless of their usage in the code, or forget to specify all dependencies in the setup.py or requirements.txt file. PyPI allows direct URL links using the ``$<$Dep Name$>$ @ $<$URL$>$" format. PyPI also uses the pip-freeze command to output installed package versions which can be used for locking dependencies, similar to package-lock in \npm. \\
\indent{It is worthwhile mentioning that while other ecosystems can differ in the type of smells they can experience, the severity of the smells might also be different. A good example is the difference in the way the JavaScript and Python package managers handle dependency conflicts. While \npm handles conflicts by installing multiple versions of a transitive dependency, which are each nested inside their respective direct dependencies \cite{npminstall}, PyPI will either outright reject installing incompatible Python packages or break another dependency by installing an incompatible transitive dependency \cite{pipguide}. This means that while using pinned dependencies can cause security issues in both ecosystems (due to not receiving security updates), in the case of Python, it can also break dependencies, while for \npm, it will only increase the installation size.}

\color{black}

\section{Implications}
\label{sec:Implications}
We present actionable implications for developers, package maintainers, researchers, and educators.

\noindent\textbf{Implications for Developers and Maintainers:}
The prevalence of dependency smells (RQ1) and their increase (Section~\ref{sec:Discussion}) implies that current dependency management practices are inadequate. \rev{We found that 80\% of the projects are infected with two or more distinct smells.} \rev{Many of the reasons for introducing smells in RQ3 (Section~\ref{sec:Results}), such as R2, R3, R4, R6, R7, R8, R9 and R10 can be fully or partially addressed if developers regard them as a priority, perhaps by opting to use SemVer rather than the alternative. For example, some developers are against a centralized repository altogether (R7) so they opt to use direct URLs, But these external links can bring about many security and stability issues (Section~\ref{sec:Dependency smells}). They also cite a mistrust that stems from historical practices but not necessarily from the package at hand (R2).} It would be beneficial for developers to spend more effort in dependency maintenance, but such maintenance requires proper guidance to be efficient. Developers can use our open source tool \cite{dependencysniffer} to quickly scan their projects and identify dependency smells as ``hot-spots" requiring prioritized maintenance. Additionally, developers could add a ``dependency maintenance" task to the code review checklist, every time a dependency is added or modified in a pull request.

The number of dependency smells in a package (given by a tool such as DependencySniffer) can also be used as a metric to evaluate and compare package quality, especially in cases where the functionality is similar. When faced with a choice between two similar packages, selecting one with more dependency smells may propagate maintenance issues to our own project in the future. \rev{It also affects the ecosystem as more projects along a dependency chain could be infected by dependency smells.} This can be accomplished by displaying ``clean" or ``infected" badges in the project's page on \npm. Badges are a great way to provide status information for the package and they can be displayed on GitHub or the \npm registry. In fact, the David DM tool attempts a similar approach by analyzing project dependencies and providing ``up to date" or ``out of date" badges for the project \cite{daviddm}. \rev{Additionally, IDEs can be a good platform for warning developers about smells. For example, the JetBrains IDE currently issues warnings for missing dependencies, but not the other smells mentioned in this paper \cite{jetbrainsinspection}.}

Developers agree that dependency smells are harmful (RQ2) but some key reasons for reluctantly resorting to dependency smells are rooted in the dependencies themselves (RQ3). \rev{Some of the reasons for introducing smells in RQ3 (Section~\ref{sec:Results}), such as R1, R2 and R5 need to be addressed by the maintainers. As such, package maintainers play a very important role in reducing the spread of dependency smells in an ecosystem. For example, developers mentioned how unexpected breaking changes (R1) forced them to pin a dependency for which they previously adhered to SemVer, or how a maintainer's apathy towards releasing requested fixes (R5) forced them away from the official package towards custom URLs. They also state how they sometimes resort to alternatives because they do not trust that SemVer is correctly respected by the maintainers (R2). Thus, proper SemVer compliance by maintainers can alleviate some of the key cited reasons for turning to dependency smells.} The Greenkeeper tool attempts to proactively warn developers when a dependency update breaks their code by running CI tests each time a dependency is updated~\cite{greenkeeper}. Package maintainers can also benefit from this by running a select number of tests from their dependents' projects before releasing a new version of the package. This can be further encouraged at the ecosystem level by labeling packages that frequently violate SemVer in their releases.

\noindent\textbf{Implications for Researchers and Educators:}
Dependency smells occur in a very considerable portion of the projects in our dataset (RQ1). However, there are yet other research aspects which should be explored. One possibility is mapping a smell-inducing commit to its fixing commit to determine the lifetime of dependency smells. Also, observing how smell type or project characteristics influence the lifetime of smells in a project will help in better understanding the root cause of dependency smells. \rev{Additionally, it is worthwhile to study which smell instances have a considerable impact on project maintenance based on revision history, issue tracking records and other project data to better understand the impact of dependency smells on software evolution.}

It is not rare to see that developers will introduce smells to some of their dependencies, while others remain clean. This poses a very important research question: do some packages have certain characteristics or behaviours that incline their dependents to introduce smells when depending on them? what causes certain packages to be more trustworthy than others? One approach would be to train a machine learning model using package features such as age in an attempt to predict how others will trust that package. \rev{Another example would be to use ``wisdom of the crowds" by analyzing how others depend on a package and recommending the most agreed-upon dependency approach as the golden standard for that package\cite{decan2019}}. This can shed more light on the reasons why dependency smells are introduced and whether all packages are equally responsible for the presence of smells in an ecosystem.

Our catalog of dependency smells contains full descriptions and consequences for seven \npm dependency smells (Section~\ref{sec:Dependency smells}). This catalog is backed up by empirical evidence regarding their prevalence (RQ1), and practitioner validations regarding their negative impact (RQ2). It is thus a great reference guide for educators looking to teach best practices in dependency management. While analyzing survey responses from 41 practitioners (RQ2), no single developer was aware of every smell, nor were they aware of every possible consequence for the smells. Teaching this collective knowledge to future developers is a good way to battle the increase in dependency smell accumulation (Section~\ref{sec:Discussion}).

\section{Related Work}
\label{sec:Related Work}
While we could not find any research works that specifically target dependency smells, the issues surrounding dependency management in software ecosystems is a well known and actively explored topic. There are also some studies which look at other types of configuration smells.

\noindent\textbf{Dependency management issues:}
Being too restrictive in updating dependencies will prevent developers from receiving the latest security and bug fixes, and as observed by Cox et al., systems with outdated dependencies are four times more likely to have security vulnerabilities \cite{cox2015measuring}. Derr et al. also found that almost 98\% of the actively used android libraries that they investigated have a security vulnerability that can be fixed by simply updating the package versions \cite{derr2017keep}. The result of this study hints that developers are sometimes aware of dependency issues, but they choose to ignore it because addressing them can be a lot of work. However, a survey by Kula et al. showed that up to 69\% of developers are unaware of the fact that they have a vulnerable dependency in their project \cite{kula2018}. The work of Decan et al. analyzes security reports for \npm packages and found that more than 40\% of releases depending on a vulnerable package could easily be protected if they were less restrictive in their updates \cite{decan2018impact}. A study by Zimmermann et al. looks at the potential of both packages and package maintainers in \npm to affect larger parts of the ecosystem. They look at over five million package versions and found that up to 40\% of all packages depend on a code that has at least one publicly disclosed security vulnerability \cite{zimmermann2019small}. They also observed how particular features of the \npm, such as pinned dependencies and heavy reuse (especially in micro-packages \cite{abdalkareem2017}) exacerbate dependency issues.

Opting for a permissive approach will facilitate updates, but increases the exposure to breaking changes. Bogart et al. study breaking changes in Eclipse, CRAN, and \npm and find that compared to other ecosystems, \npm developers are more willing to perform breaking changes with the assumption that users that use SemVer will be protected \cite{bogart2016break}. Bogart et al. also conduct a survey of 2,000 developers across eighteen ecosystems where 70\% of the respondents for \npm declare that they have experienced breaking changes when attempting to build their package~\cite{values2020}. 
A study on the Maven ecosystem finds that more than 35\% of minor and more than 23\% of patch releases contain at least one breaking change \cite{raemaekers2014semantic}.
\color{black}
Semantic versioning is one of the proposed solutions to the dependency management issues. A recent work by Decan and Mens \cite{decan2019} looks at the four ecosystems of Cargo, \npm, Packagist, and RubyGems. They found that there is a promising trend towards SemVer compliance, but in the cases where SemVer is not followed, developers prefer to be restrictive rather than permissive. An empirical study by Decan et al. looks into dependency issues in \npm, CRAN, and RubyGems, focusing on the extent of dependency relationships and the issues surrounding dependency constraints\cite{decan2017empirical}. They find that developers prefer to specify a maximum threshold on their constraints, rather than a minimum one. They also mention the co-installability issues with strict (aka pinned) dependencies. Another recent work looks at how dependencies adopt SemVer and how they change their approach over time. \rev{Dietrich et al.} were unable to find a large-scale adoption of SemVer and there is evidence of both flexible and restrictive approaches to update\cite{dietrich2019dependency}. These studies are all solely based on \npm packages and do not consider applications, which can have different approaches in managing dependencies. They also do not mention issues beyond permissive or restrictive constraints, or quantify how these issues evolve over time. 

To the best of our knowledge, our study is the first work that specifically focuses on cataloging, quantifying, and understanding dependency smells.
 
\noindent\textbf{Configuration smells:}
Infrastructure as Code (IaC) is another domain where system configuration is specified through code. While configuration code is different than source code, it is subject to similar issues of maintainability and complexity. Sharma et al. \cite{sharma2016does} analyzed more than 4,600 Puppet repositories and proposed a catalog of configuration smells that violate best practices. 
Configuration scripts can also contain smells related to security vulnerabilities. These security smells are reoccurring patterns of configuration code snippets which can lead to security breaches. Rahman et al. conduct an empirical study on more than 1,700 IaC scripts \cite{rahman2019seven} to identify seven security smells. They then developed a detection tool and collected instances of those smells from 293 open-source repositories and submitted bug reports for a random subset of the instances. They found that security smells have a median lifetime of 20 months while some can persist for as long as 98 months. 
Since the package.json has a relatively well-defined structure and functionality, it is possible to automatically parse it to detect dependency smells. 
\rev{Semantic versioning guidelines can also be an effective solution in Infrastructure as Code. Opdebeeck et al. study 70,000 version increments in the Ansible infrastructure and discover that Ansible role developers generally abide by semantic versioning guidelines \cite{opdebeeck2020does}.}

\section{Threats to Validity}
\label{sec:Threats to Validity}
In this section, we discuss the threats to validity of our study.\\
\noindent\textbf{Threats to construct validity:} 
Threats to construct validity refers to the concern between the theory and the results of the study.
The smells catalog in this study is neither exhaustive nor complete. There might be other dependency smells in the \npm ecosystem that our study did not consider. In addition, many of our dependency smells are observed through the package.json file. However, it is possible for developers to manually install dependencies and not include them in package.json. While we have studied such cases under ``Unused dependencies", they could also be subject to other smells such as "Pinned dependency", since a manually installed package may not be automatically updated. However, since package.json is the official dependency configuration file for \npm, ad-hoc alternatives are rare and discouraged. 
Also, while investigating the reasons for the introduction of smells (RQ3), we emailed the developers in charge of the smell-inducing commit. In reality, there may be other sources, such as other developers involved in group discussions, that may have also influenced the change. We believe the developer in charge of the commit would be the most-informed member of the team on this change and in fact, all respondents were able to elaborate on the commit.
\color{black}

\noindent\textbf{Threats to internal validity:} 
Threats to internal validity refers to the concerns that are internal to the study such as experimenter bias and errors.
In our work, we used the depcheck~\cite{depcheck2019} tool to extract the missing and unused dependency smells in the studied projects. Hence, we are limited by the accuracy of this tool. To mitigate the threats related to using the depcheck tool, we randomly selected five projects from our dataset and fed them to the depcheck tool. We then manually cross-checked the output of this tool. We found that in all cases, the tool produced the correct results for missing dependencies, as we were able to navigate to the file that uses a dependency not specified in the package.json. Validating unused dependencies can be much harder since there are different ways to use a dependency inside a code. While the results proved accurate in our manual checks, this is one area where depcheck users have experienced false positives. 

\rev{Similar to the process used for empirically studying the smells, our proposed tool uses regular expressions to detect the first four dependency smells. To validate the accuracy of the regular expressions and the parser as a whole, we manually examined a sample of 10\% of the cases and we observed no false positives or false negatives. However, our manual validation is still limited to 10\% of the dataset. Additionally, we used the tool to identify smells and emailed developers for clarification (RQ3), and all respondents verified the existence of the smells. These results make us confident in our parsing process and our regular expressions.}

\rev{Our initial survey gathered responses from 12  developers. These developers were a convenience sample which were known by the authors in Canada and Brazil which is not representative of all JavaScript developers. However, this survey was conducted to augment the initial understanding of the smells. The initial definitions are observed by studying violations of \npm recommendations, violations of SemVer guidelines and by studying the discussions surrounding the package.json file.}
In addition, to extract the reasons for introducing a dependency smell, we manually analyzed the free-text responses provided by the developers. Since this human activity and can be subject to human bias, we perform a thematic analysis with two independent coders where we have the first two authors independently analyze and categorize them.

\color{black}

\noindent\textbf{Threats to external validity:}
Threats to external validity concerns the generalization of our findings.
Our study is based solely on JavaScript projects, therefore our findings may not hold for projects written in other programming languages. However, our research methodology of defining and studying dependency smells can be easily replicated for other programming languages such as Python. Secondly, the datasets that we used in our work only represent open source projects hosted on GitHub that may not reflect proprietary projects. Furthermore, we examine the official dependency manager for JavaScript, \npm. Hence, our results may not be fully generalized to other dependency managers such as yarn. However, yarn uses the same package.json file to evaluate dependencies and uses the same \npm registry to download them \cite{yarndoc2020}. The only difference related to our work is cases where projects have a yarn-lock file instead of a package-lock. That is why we considered projects with yarn.lock to be free of the no package-lock smell.
To understand why dependency smells are introduced in projects, we surveyed 28 developers. Although we believe this to be a sufficient number of developers for our analysis, our results may not represent the opinion of all JavaScript developers. Also, asking a different sample of developers may result in a different set of reasons for introducing dependency smells. To mitigate the threat, we contacted practitioners from different studied projects and their backgrounds show that they are experienced JavaScript developers. 

\noindent\textbf{Threats to conclusion validity:} 
Threats to conclusion validity concern the relation between the experiment and the conclusions.
The empirical study of dependency smells is based on historical data in 1146 JavaScript projects, but investigating their impact and the reasons for their introduction is conducted using surveys. Therefore, the conclusions drawn are based on the type and number of respondents. The response rate for the emails we sent out to developers to understand the reasons for introducing the smells was 28\%. While this is significantly larger than the typical rate in questionnaire-based software engineering surveys \cite{singer2008software}, having a larger portion of respondents may reveal new reasons or change the priority of current reasons.
Additionally, we argue that the prevalence of dependency smells in RQ1 implies a lack of attention to dependency maintenance. The dependency smells in this paper focus on how dependencies are used, but do not consider other aspects of dependency maintenance, such as what dependencies are selected. Therefore, dependency smells alone are not enough to compare dependency maintenance across projects. Hence, this study does not claim to measure dependency maintenance.
\color{black}

\section{Conclusion}
\label{sec:Conclusion}
Our objective was to catalog, quantify, and understand dependency smells in the \npm ecosystem. We conducted an empirical study on a dataset of 1146 active JavaScript projects to identify which smells are more common and how they accumulate over time. We also consulted practitioners to quantify the consequences associated with each smell and understand why they are introduced. 

We define seven dependency smells in \npm. Our findings reveal that these smells are prevalent. \rev{While not all smells occur at a large degree, 80\% of the projects are infected with two or more distinct smells.} In our practitioner surveys, we found that practitioners recognized the multitude of security problems, bugs, dependency breakages, and other maintenance issues brought about by dependency smells. These smells are generally introduced ass developers react to dependency misbehavior and the shortcomings of the \npm ecosystem. We also observed that dependency smells are addressed/fixed, but most dependency smells are introduced more frequently than old smells are fixed, causing an accumulation of smells over time.

Since we had to analyze a large number of projects, commits, and smells, we built a tool (DependencySniffer) for our analyses. Our prototype tool can be used to analyze any JavaScript project and detect dependency smells. The tool is open source and accessible to everyone \cite{dependencysniffer}.
\\

\section*{Acknowledgements}
\label{sec: Acknowledgements}
The authors are grateful to the respondents who dedicated their time to participate in our surveys. The involvement of human participants in this study was approved by Concordia University’s Faculty Research \& Ethics Advisory Committee (summary protocol form number: 30004729).
\color{black}

\bibliographystyle{abbrv}
\balance
\bibliography{abbas-base}

\begin{thebibliography}{10}

\bibitem{abdalkareem2017}
R.~Abdalkareem, O.~Nourry, S.~Wehaibi, S.~Mujahid, and E.~Shihab.
\newblock Why do developers use trivial packages? an empirical case study on
  npm.
\newblock In {\em Proceedings of the 2017 11th Joint Meeting on Foundations of
  Software Engineering}, pages 385--395. ACM, 2017.

\bibitem{artho2012software}
C.~Artho, K.~Suzaki, R.~Di~Cosmo, R.~Treinen, and S.~Zacchiroli.
\newblock Why do software packages conflict?
\newblock In {\em Proceedings of the 9th IEEE Working Conference on Mining
  Software Repositories}, pages 141--150. IEEE Press, 2012.

\bibitem{values2020}
C.~Bogart, A.~Filippova, C.~Kästner, and J.~Herbsleb.
\newblock How ecosystem cultures differ: Results from a survey on values and
  practices across 18 software ecosystems.
\newblock \url{http://breakingapis.org/survey/}, October 2017.
\newblock (accessed on 10/16/2020).

\bibitem{bogart2016break}
C.~Bogart, C.~K{\"a}stner, J.~Herbsleb, and F.~Thung.
\newblock How to break an {API}: cost negotiation and community values in three
  software ecosystems.
\newblock In {\em Proceedings of the 2016 24th ACM SIGSOFT International
  Symposium on Foundations of Software Engineering}, pages 109--120. ACM, 2016.

\bibitem{burnard1991method}
P.~Burnard.
\newblock A method of analysing interview transcripts in qualitative research.
\newblock {\em Nurse education today}, 11(6):461--466, 1991.

\bibitem{underscore2014}
D.~Chatfield.
\newblock Fix the versioning · issue \#1805 · jashkenas/underscore.
\newblock \url{https://github.com/jashkenas/underscore/issues/1805}, September
  2014.
\newblock (Accessed on 10/16/2020).

\bibitem{chinthanet2019lag}
B.~Chinthanet, R.~G. Kula, T.~Ishio, A.~Ihara, and K.~Matsumoto.
\newblock On the lag of library vulnerability updates: An investigation into
  the repackage and delivery of security fixes within the npm {JavaScript}
  ecosystem.
\newblock {\em arXiv preprint arXiv:1907.03407}, 2019.

\bibitem{pipsemver}
N.~Coghlan and D.~Stufft.
\newblock Version identification and dependency specification.
\newblock \url{https://www.python.org/dev/peps/pep-0440}, February 2021.
\newblock (accessed on 3/20/2021).

\bibitem{cox2015measuring}
J.~Cox, E.~Bouwers, M.~Van~Eekelen, and J.~Visser.
\newblock Measuring dependency freshness in software systems.
\newblock In {\em 2015 IEEE/ACM 37th IEEE International Conference on Software
  Engineering}, volume~2, pages 109--118. IEEE, 2015.

\bibitem{decan2019}
A.~Decan and T.~Mens.
\newblock What do package dependencies tell us about semantic versioning?
\newblock {\em IEEE Transactions on Software Engineering}, 2019.

\bibitem{decan2017empirical}
A.~Decan, T.~Mens, and M.~Claes.
\newblock An empirical comparison of dependency issues in {OSS} packaging
  ecosystems.
\newblock In {\em 2017 IEEE 24th International Conference on Software Analysis,
  Evolution and Reengineering (SANER)}, pages 2--12. IEEE, 2017.

\bibitem{decan2018evolution}
A.~Decan, T.~Mens, and E.~Constantinou.
\newblock On the evolution of technical lag in the npm package dependency
  network.
\newblock In {\em 2018 IEEE International Conference on Software Maintenance
  and Evolution (ICSME)}, pages 404--414. IEEE, 2018.

\bibitem{decan2018impact}
A.~Decan, T.~Mens, and E.~Constantinou.
\newblock On the impact of security vulnerabilities in the npm package
  dependency network.
\newblock In {\em 2018 IEEE/ACM 15th International Conference on Mining
  Software Repositories (MSR)}, pages 181--191. IEEE, 2018.

\bibitem{decan2019empirical}
A.~Decan, T.~Mens, and P.~Grosjean.
\newblock An empirical comparison of dependency network evolution in seven
  software packaging ecosystems.
\newblock {\em Empirical Software Engineering}, 24(1):381--416, 2019.

\bibitem{derr2017keep}
E.~Derr, S.~Bugiel, S.~Fahl, Y.~Acar, and M.~Backes.
\newblock Keep me updated: An empirical study of third-party library
  updatability on {Android}.
\newblock In {\em Proceedings of the 2017 ACM SIGSAC Conference on Computer and
  Communications Security}, pages 2187--2200. ACM, 2017.

\bibitem{dietrich2019dependency}
J.~Dietrich, D.~J. Pearce, J.~Stringer, A.~Tahir, and K.~Blincoe.
\newblock Dependency versioning in the wild.
\newblock In {\em Proceedings of the 16th International Conference on Mining
  Software Repositories}, pages 349--359. IEEE Press, 2019.

\bibitem{Fincher_2005OpenCardSort}
S.~Fincher and J.~Tenenberg.
\newblock Making sense of card sorting data.
\newblock {\em Expert Systems}, 22(3):89--93, 2005.

\bibitem{fontana2016antipattern}
F.~A. Fontana, J.~Dietrich, B.~Walter, A.~Yamashita, and M.~Zanoni.
\newblock Antipattern and code smell false positives: Preliminary
  conceptualization and classification.
\newblock In {\em 2016 IEEE 23rd international conference on software analysis,
  evolution, and reengineering (SANER)}, volume~1, pages 609--613. IEEE, 2016.

\bibitem{TheState18online}
GitHub.
\newblock The state of the octoverse | the state of the octoverse celebrates a
  year of building across teams, time zones, and millions of merged pull
  requests.
\newblock \url{https://octoverse.github.com/}, 2019.

\bibitem{goodman1961snowball}
L.~A. Goodman.
\newblock Snowball sampling.
\newblock {\em The annals of mathematical statistics}, pages 148--170, 1961.

\bibitem{Ghtorrent2019}
G.~Gousios.
\newblock The {GHTorrent} project.
\newblock \url{https://ghtorrent.org/}, Mar 2019.
\newblock (accessed on 10/16/2020).

\bibitem{dependencysniffer}
A.~Javan~Jafari.
\newblock {Dependency Sniffer}.
\newblock \url{https://github.com/abbasjavan/DependencySniffer}, September
  2020.

\bibitem{replication}
A.~Javan~Jafari, D.~Elias~Costa, R.~Abdalkareem, E.~Shihab, and N.~Tsantalis.
\newblock Replication package for dependency smells in {JavaScript} projects.
\newblock \url{https://doi.org/10.5281/zenodo.4701497}, March 2021.

\bibitem{jetbrainsinspection}
JetBrains.
\newblock Code inspections in {JavaScript} and {TypeScript}.
\newblock
  \url{https://www.jetbrains.com/help/phpstorm/code-inspections-in-javascript-and-typescript.html#Imports_and_dependencies},
  March 2021.
\newblock (accessed on 26/04/2021).

\bibitem{Kalliamvakou:14:MiningGithub}
E.~Kalliamvakou, G.~Gousios, K.~Blincoe, L.~Singer, D.~M. German, and
  D.~Damian.
\newblock The promises and perils of mining {GitHub}.
\newblock In {\em Proceedings of the 11th Working Conference on Mining Software
  Repositories}, MSR 2014, page 92–101, New York, NY, USA, 2014. Association
  for Computing Machinery.

\bibitem{kula2018}
R.~G. Kula, D.~M. German, A.~Ouni, T.~Ishio, and K.~Inoue.
\newblock Do developers update their library dependencies?
\newblock {\em Empirical Software Engineering}, 23(1):384--417, 2018.

\bibitem{greenkeeper}
J.~Lehnardt and S.~Haas.
\newblock Greenkeeper.
\newblock \url{https://greenkeeper.io/docs.html}, September 2020.
\newblock (accessed on 10/14/2020).

\bibitem{depcheck2019}
J.~Li and D.~Lukic.
\newblock Depcheck: Check your npm module for unused dependencies.
\newblock \url{https://github.com/depcheck/depcheck}, October 2019.
\newblock (accessed on 10/16/2020).

\bibitem{libraries.io2020}
Libraries.io.
\newblock npm on libraries.io.
\newblock \url{https://libraries.io/npm}, September 2020.
\newblock (accessed on September, 2020).

\bibitem{lim1994effects}
W.~C. Lim.
\newblock Effects of reuse on quality, productivity, and economics.
\newblock {\em IEEE software}, 11(5):23--30, 1994.

\bibitem{leftpad2018}
F.~MacDonald.
\newblock How a programmer nearly broke the internet by deleting just 11 lines
  of code.
\newblock \url{https://www.sciencealert.com/how-a-programmer-almost-broke-the\-
  internet-by-deleting-11-lines-of-code}, September 2018.

\bibitem{mohagheghi2004empirical}
P.~Mohagheghi, R.~Conradi, O.~M. Killi, and H.~Schwarz.
\newblock An empirical study of software reuse vs. defect-density and
  stability.
\newblock In {\em Proceedings of the 26th international conference on software
  engineering}, pages 282--292. IEEE Computer Society, 2004.

\bibitem{npmtypo2017}
npm.
\newblock The npm blog — new package moniker rules.
\newblock
  \url{https://blog.npmjs.org/post/168978377570/new-package-moniker-rules},
  December 2017.
\newblock (accessed on 10/16/2020).

\bibitem{npmsemver}
npm.
\newblock About semantic versioning.
\newblock \url{https://docs.npmjs.com/about-semantic-versioning}, February
  2021.

\bibitem{npminstall}
npm.
\newblock npm install documentation.
\newblock \url{https://docs.npmjs.com/cli/v6/commands/npm-install}, February
  2021.

\bibitem{npmpkglock2019}
npm Documentation.
\newblock npm-package-lock.json | npm documentation.
\newblock \url{https://docs.npmjs.com/files/package-lock.json}, October 2019.
\newblock (accessed on 10/16/2020).

\bibitem{npmpkg2019}
npm Documentation.
\newblock npm-package.json | npm documentation.
\newblock \url{https://docs.npmjs.com/files/package.json}, October 2019.

\bibitem{opdebeeck2020does}
R.~Opdebeeck, A.~Zerouali, C.~Vel{\'a}zquez-Rodr{\'\i}guez, and C.~De~Roover.
\newblock Does infrastructure as code adhere to semantic versioning? an
  analysis of ansible role evolution.
\newblock In {\em 2020 IEEE 20th International Working Conference on Source
  Code Analysis and Manipulation (SCAM)}, pages 238--248. IEEE, 2020.

\bibitem{nodesource}
T.~Oxley.
\newblock Semver: Tilde and caret.
\newblock \url{https://nodesource.com/blog/semver-tilde-and-caret}, September
  2014.
\newblock (accessed on 3/22/2021).

\bibitem{semver2019}
T.~Preston-Werner.
\newblock Semantic versioning 2.0.0 | semantic versioning.
\newblock \url{https://semver.org/}, March 2019.
\newblock (accessed on 10/16/2020).

\bibitem{pipguide}
PyPA.
\newblock pip documentation.
\newblock \url{https://pip.pypa.io/en/latest/user_guide/}, February 2021.
\newblock (accessed on 2/20/2021).

\bibitem{raemaekers2014semantic}
S.~Raemaekers, A.~Van~Deursen, and J.~Visser.
\newblock Semantic versioning versus breaking changes: A study of the maven
  repository.
\newblock In {\em 2014 IEEE 14th International Working Conference on Source
  Code Analysis and Manipulation}, pages 215--224. IEEE, 2014.

\bibitem{rahman2019seven}
A.~Rahman, C.~Parnin, and L.~Williams.
\newblock The seven sins: security smells in infrastructure as code scripts.
\newblock In {\em Proceedings of the 41st International Conference on Software
  Engineering}, pages 164--175. IEEE Press, 2019.

\bibitem{cargowild2018}
Rust-lang.
\newblock Rust edition guide.
\newblock
  \url{https://doc.rust-lang.org/edition-guide/rust-2018/cargo-and-crates-io/crates-io-disallows-wildcard-dependencies.html},
  December 2018.
\newblock (accessed on September, 2020).

\bibitem{sharma2016does}
T.~Sharma, M.~Fragkoulis, and D.~Spinellis.
\newblock Does your configuration code smell?
\newblock In {\em 2016 IEEE/ACM 13th Working Conference on Mining Software
  Repositories (MSR)}, pages 189--200. IEEE, 2016.

\bibitem{daviddm}
A.~Shaw.
\newblock David-dm.
\newblock \url{https://david-dm.org}, September 2020.

\bibitem{singer2008software}
J.~Singer, S.~E. Sim, and T.~C. Lethbridge.
\newblock Software engineering data collection for field studies.
\newblock In {\em Guide to Advanced Empirical Software Engineering}, pages
  9--34. Springer, 2008.

\bibitem{soto2020comprehensive}
C.~Soto-Valero, N.~Harrand, M.~Monperrus, and B.~Baudry.
\newblock A comprehensive study of bloated dependencies in the maven ecosystem.
\newblock {\em arXiv preprint arXiv:2001.07808}, 2020.

\bibitem{wittern2016look}
E.~Wittern, P.~Suter, and S.~Rajagopalan.
\newblock A look at the dynamics of the {JavaScript} package ecosystem.
\newblock In {\em 2016 IEEE/ACM 13th Working Conference on Mining Software
  Repositories (MSR)}, pages 351--361. IEEE, 2016.

\bibitem{yarndoc2020}
Yarn.
\newblock Documentation | yarn.
\newblock \url{https://classic.yarnpkg.com/en/docs/}, 2020 2020.
\newblock (accessed on 10/16/2020).

\bibitem{zimmermann2019small}
M.~Zimmermann, C.-A. Staicu, C.~Tenny, and M.~Pradel.
\newblock Small world with high risks: A study of security threats in the npm
  ecosystem.
\newblock {\em arXiv preprint arXiv:1902.09217 (To appear in the 28th USENIX
  Security Symposium)}, 2019.

\end{thebibliography}

\vspace{-20 mm}
\begin{IEEEbiography}[{\includegraphics[width=1in,height=1.25in,clip,keepaspectratio]{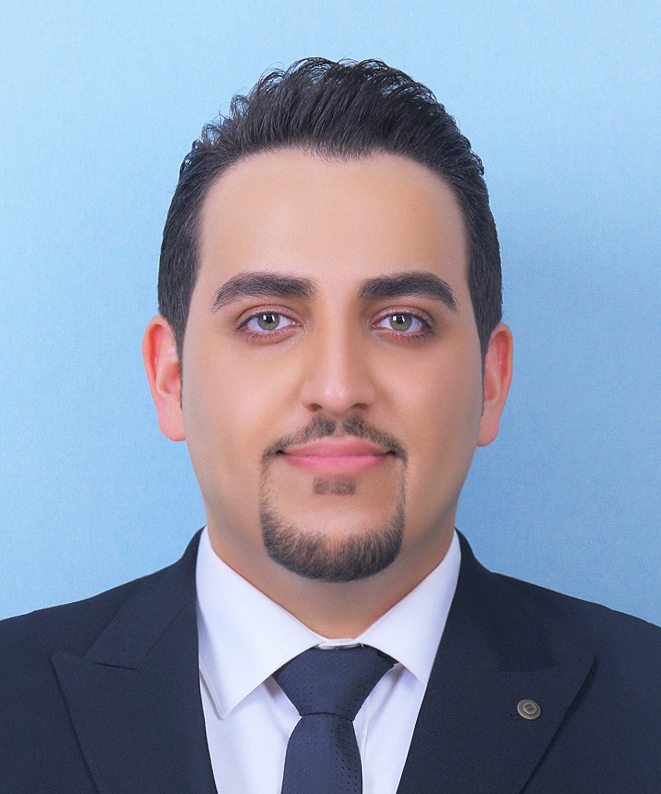}}]{Abbas Javan Jafari} 
is a PhD Student at the DAS Lab, in the Department of Computer Science and Software Engineering at Concordia University, Canada. His research interests include mining software repositories, software ecosystems, empirical software engineering and software quality with an emphasis on security. 
You can find more information at http://das.encs.concordia.ca/members/abbas-javan/.
\end{IEEEbiography}
\vspace{-20 mm}
\begin{IEEEbiography}[{\includegraphics[width=1in,height=1.25in,clip,keepaspectratio]{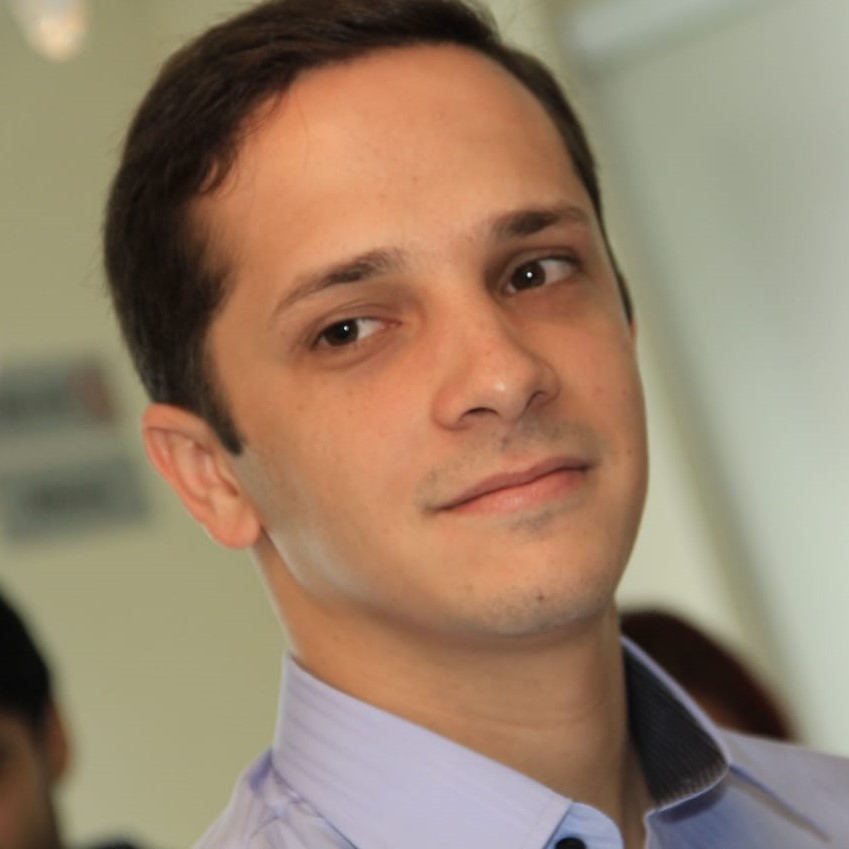}}]{Diego Elias Costa}
	is a postdoctoral researcher at the DAS Lab, in the Department of Computer Science and Software Engineering at Concordia University. His research interests cover a wide range of software engineering and performance engineering related topics, including mining software repositories, empirical software engineering, performance testing, memory-leak detection, and adaptive data structures. 
	You can find more about him at \url{http://das.encs.concordia.ca/members/diego-costa/}.
\end{IEEEbiography}
\vspace{-20 mm}
\begin{IEEEbiography}[{\includegraphics[width=1in,height=1.25in,clip,keepaspectratio]{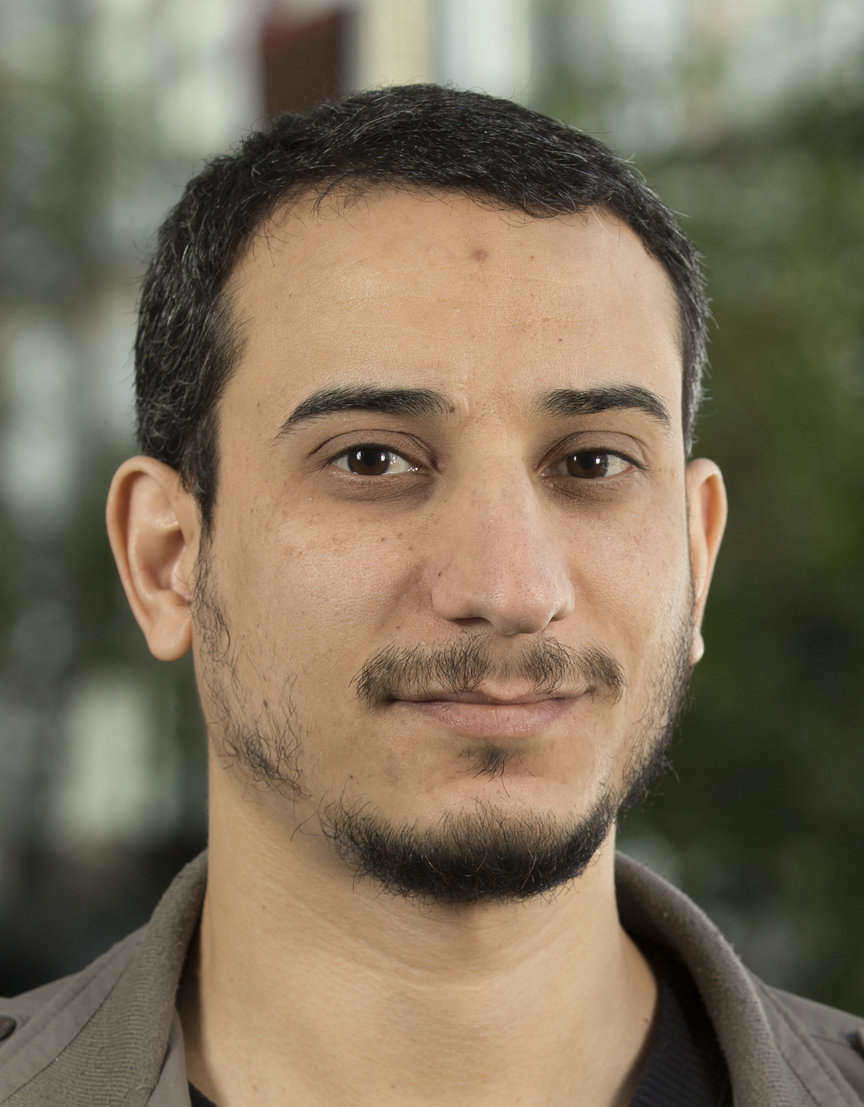}}]{Rabe Abdalkareem} is an assistant professor in the School of Computer Science at Carleton University. He obtained his Ph.D. in Computer Science and Software Engineering from Concordia University, Canada. His research investigates how the adoption of crowd-sourced knowledge affects software development and maintenance. Abdalkareem received his master’s in applied computer science from Concordia University. His work has been published at premier venues such as FSE, ICSME, and MSR and major journals such as TSE, IEEE Software, EMSE, and IST. More information can be found at \url{https://rabeabdalkareem.github.io/}
\end{IEEEbiography}
\vspace{-10 mm}
\begin{IEEEbiography}[{\includegraphics[width=1in,height=1.25in,clip,keepaspectratio]{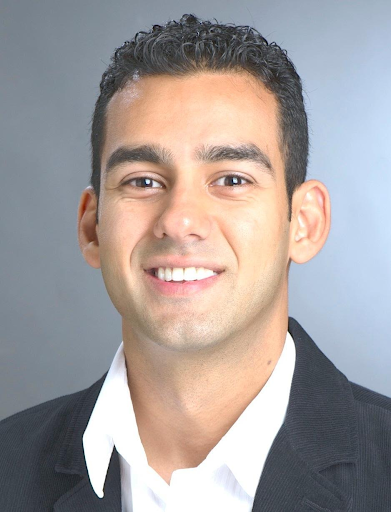}}]{Emad Shihab} is an associate professor in the Department of Computer Science and Software Engineering at Concordia University. He received his PhD from Queens University. Dr. Shihabs research interests are in Software Quality Assurance, Mining Software Repositories, Technical Debt, Mobile Applications and Software Architecture. He worked as a software research intern at Research In Motion in Waterloo, Ontario and Microsoft Research in Redmond, Washington. Dr. Shihab is a member of the IEEE and ACM. More information can be found at http://das.encs.concordia.ca.
\end{IEEEbiography}
\vspace{-10 mm}
\begin{IEEEbiography}[{\includegraphics[width=1in,height=1.25in,clip,keepaspectratio]{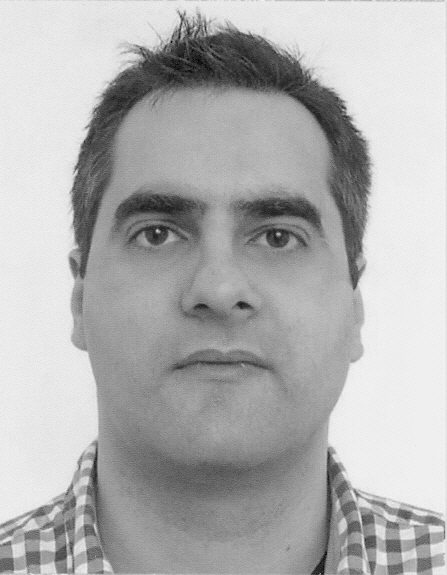}}]{Nikolaos Tsantalis} is an associate professor in the department of Computer Science and Software Engineering at Concordia University, Montreal, Canada, and holds a Concordia University Research Chair in Web Software Technologies. His research interests include software maintenance, software evolution, empirical software engineering, refactoring recommendation systems, refactoring mining, and software quality assurance. He has been awarded with two Most Influential Paper awards at SANER 2018 and SANER 2019, and two ACM SIGSOFT Distinguished Paper awards at FSE 2016 and ICSE 2017. He served as a program co-chair for various tracks in ICSME, SANER, SCAM and ICPC conferences, and serves regularly as a program committee member of international conferences in the field of software engineering, such as ASE, ICSME, MSR, SANER, ICPC, and SCAM. He is a member of the IEEE TSE review board. Finally, he is a senior member of the IEEE and the ACM, and holds a license from the Association of Professional Engineers of Ontario.
\end{IEEEbiography}




\end{document}